\documentclass[prb,showpacs,twocolumn,notitlepage,final,floatfix]{revtex4}

\usepackage{natbib}
\usepackage{graphicx}
\usepackage{amsmath,amssymb,amsbsy,latexsym}
\usepackage{bbm}

\graphicspath{{../images-cmyk/}}

\newcommand{\eqnref}[1]{(\ref{#1})}
\newcommand{\figref}[1]{\ref{#1}}
\newcommand{\secref}[1]{\ref{#1}}

\newcommand{\viscosity}{\eta}
\newcommand{\tension}{\gamma}
\newcommand{\stress}{\sigma}
\newcommand{\curv}{\kappa}
\newcommand{\Capillary}{\textit{Ca}}
\newcommand{\Bond}{\textit{Bo}}
\newcommand{\Reynolds}{\textit{Re}}

\newcommand{\bfe}{\mathbf{e}}

\newcommand{\bfN}{\mathbf{N}}
\newcommand{\bfT}{\mathbf{T}}
\newcommand{\bft}{\mathbf{t}}
\newcommand{\bfv}{\mathbf{v}}
\newcommand{\bfx}{\mathbf{x}}
\newcommand{\bfnu}{\boldsymbol{\nu}}
\newcommand{\calP}{\mathcal{P}}
\newcommand{\calF}{\mathcal{F}}
\newcommand{\vecp}{\vec{p}}
\newcommand{\vecu}{\vec{u}}
\newcommand{\vecv}{\vec{v}}
\newcommand{\vecr}{\vec{r}}
\newcommand{\vecs}{\vec{s}}
\newcommand{\formal}{\partial}
\newcommand{\variat}{\delta}
\newcommand{\spacedim}{D}
\newcommand{\Real}{\mathbbm{R}}
\newcommand{\bfnabla}{\pmb\nabla}
\newcommand{\scalarmult}{\!\boldsymbol{\cdot}\!}

\newcommand{\etal}{\textit{et.\,al.}}

\begin{document}

\title[Computing stationary free-surface shapes in microfluidics]%
{Computing stationary free-surface shapes in microfluidics}

\author{Michael Schindler}
\author{Peter Talkner}
\author{Peter H\"anggi}
\affiliation{Institut f\"ur Physik,
Universit\"atsstra{\ss}e 1, 86159 Augsburg, Germany}

\date{\today}


\begin{abstract}
A finite-element algorithm for computing free-surface flows driven by arbitrary
body forces is presented. The algorithm is primarily designed for the
microfluidic parameter range where (i) the Reynolds number is small and (ii)
force-driven pressure and flow fields compete with the surface tension for the
shape of a stationary free surface. The free surface shape is represented by the
boundaries of finite elements that move according to the stress applied by the
adjacent fluid. Additionally, the surface tends to minimize its free energy and
by that adapts its curvature to balance the normal stress at the surface. The
numerical approach consists of the iteration of two alternating steps: The
solution of a fluidic problem in a prescribed domain with slip boundary
conditions at the free surface and a consecutive update of the domain driven by
the previously determined pressure and velocity fields. For a Stokes problem the
first step is linear, whereas the second step involves the nonlinear
free-surface boundary condition. This algorithm is justified both by physical
and mathematical arguments. It is tested in two dimensions for two cases that
can be solved analytically. The magnitude of the errors is discussed in
dependence on the approximation order of the finite elements and on a step-width
parameter of the algorithm. Moreover, the algorithm is shown to be robust in the
sense that convergence is reached also from initial forms that strongly deviate
from the final shape. The presented algorithm does not require a remeshing of
the used grid at the boundary. This advantage is achieved by a built-in
mechanism that causes a smooth change from the behavior of a free surface to
that of a rubber band if the boundary mesh becomes irregular. As a side effect,
the element sides building up the free surface in two dimensions all approach
equal lengths. The presented variational derivation of the boundary condition
corroborates the numerical finding that a second-order approximation of the
velocity also necessitates a second-order approximation for the free surface
discretization.
\end{abstract}

\maketitle

\section{Introduction}

In the past decade the development of so-called ``labs-on-a-chip''
\cite{FigPin00,StoStrAjd04} has led to an increased interest in
microfluidics,\cite{PolHay01,SquQua05,Thiele03} i.e.\ in the field of
hydrodynamics with characteristic length scales of less than a millimeter. These
flows are characterized by small \emph{Reynolds numbers} and consequently
governed by the Stokes equations. In the case of prescribed fluid domains with
no-slip boundary conditions standard numerical methods exist for computing their
solutions.\cite{DevFisMun02,ZieTay00}

Recently, various experimental techniques\cite{RamMorGre98,GutRatKel04} have
been developed to induce and control flows in fluids which sit on a substrate
without being confined by lateral and covering
walls.\cite{WixStrGau04,GutMulHab05} In the experiments the fluid is kept
together by its surface tensions both at the substrate and at the fluid--air
interface. The stationary form that is assumed by the fluid--air interface is
not given a priori. It results from an interplay of the internal streaming
pattern, the internal pressure distribution and the surface tension. On the
other hand, the form of the interface acts back on the flow. This mutual
interaction of form and flow renders free boundary value problems fascinating
but difficult. The relative importance of viscous flow and pressure, each
compared to the influence of the surface tension, can be quantified by two
dimensionless numbers, the \emph{capillary number} and a generalized \emph{Bond
number}, respectively.

In the present work we consider a small water droplet (around 50\,nl or less).
The droplet sits on a flat substrate and is mechanically agitated by a body
force.\cite{SAW} Inside the droplet this body force then causes stationary
pressure and flow fields which can lead to a significant deformation for the
free surface. Sufficiently strong body forces may lead to the motion of the
entire droplet, but this situation will not be considered here. The values of
all, the Reynolds number, the capillary number, and the Bond number are assumed
to range from zero up to unity. This corresponds to experimentally relevant
situations.\cite{SriStrSch05,WixStrGau04,GutMulHab05}

Several numerical approaches for determining free surface shapes have been
proposed in the past. The suitability of the approaches depends on the size of
the system, typical velocities, and the material properties, as well as on the
resulting deformation of the fluid domain. They can roughly be classified into
two groups: Either a fixed grid and a function describing the position of the
free surface is used, or the computational mesh is moved together with the fluid
domain, yielding a sharp surface representation by elements' boundaries.

An established method of the first kind is the \emph{continuum method} proposed
by~Brackbill \etal\cite{BraKotZem92} They circumvented the discretization of the
normal-stress boundary condition by introducing a body force density that is
concentrated near the free surface. This force density accounts for the effect
of surface tension. We have tested this method, which is implemented in the
commercially available fluid-dynamics program~\emph{FLUENT} using a
\emph{volume-of-fluid} discretization. For a macroscopic system this method
worked fine. The method, however, fails if the system is scaled down to the
microfluidic parameter regime. In a simple test example we found that
approximation errors of the free-surface boundary condition contributed to the
force balance in the Navier--Stokes equations and were amplified in an
uncontrolled manner. This typically gave rise to a spurious velocity field. It
even occurred when we started the iteration with the known solution. Problems
with this method have also been reported by Renardy~\& Renardy\cite{RenRen02}
and by Popinet~\& Zaleski.\cite{PopZal99} Lafaurie~\etal\cite{LafNarScaZalZan94}
find the spurious velocities to be of the order surface-tension/viscosity which
is the dominant velocity scale for microfluidic systems. Thus, the existing
continuum method appears to be inappropriate for the microfluidic parameter
regime.

Another approach of the first kind has recently been proposed by
Smolianski.\cite{Smolianski05} He uses finite elements and a level-set
description for the free surface and calculates curvatures by derivatives of the
distance-function. He still encounters spurious velocity fields proportional to
the ratio surface-tension/viscosity.

Methods of the second kind, representing the free surface by a sharp interface
are expected to work better in the microfluidic parameter regime. Algorithms in
this class are often referred to as ``moving mesh'' or ``ALE'' methods and
generally require more involved techniques, keeping the computational mesh
feasible and not too distorted.

A technique of the second kind that has successfully been employed for
tension-dominated free-surface problems is the boundary-element
method.\cite{Pozrikidis92,ZinRotDav97} The dimensionality of the equations is
reduced to the dimensionality of the surface which provides the basis for an
efficient implementation. Unfortunately, this reduction can only be performed
for Stokes equations with conservative body forces, which can be absorbed into
the pressure term. In the present investigation we allow for non-conservative
body forces which are of particular experimental
relevance.\cite{WixStrGau04,GutRatKel04}

Pioneering works for the finite-element implementation of the full free-surface
problem were published by Scriven and coworkers.\cite{SaiScr81,KisScr83} They
used spines to parameterize the movements of the computational mesh in coating
flow and implemented Newton's method for a Galerkin approximation scheme. This
work was later continued under the designation ``total linearization method'' by
Cuvelier and coworkers.\cite{CuvSch90,CuvSegSte86} Their description requires a
height function for the free-surface position, which makes it necessary to use
well-adapted coordinate systems like polar cylindrical or spherical ones. It
must be known in advance if a free surface will overhang.

In the present paper we extend the works of Scriven and Cuvelier to arbitrary
surface geometries. In our description, the parameterization of the free surface
is given directly by the finite-elements' boundary parameterization. Thus,
neither spines nor a height function are needed. To properly account for
intrinsic curvatures of the free surface, all equations are formulated in a
fully covariant form that allows for all differential-geometric properties of
the surface. An excellent reference for this formulation are the works of
Aris\cite{Aris89} and Scriven\cite{Scriven60} where the fluidic flow inside a
curved free surface is described.

Recently, algorithms have been published that describe fully time-dependent
free-surface flows, even in three
dimensions.\cite{Bansch98,CaiSchBaeRekRaoSac00,WalGasJimKelSum05} In these works
the free surface is moved mainly due to the \emph{kinematic} boundary condition,
i.e.~it is advected passively. Concerning convergence, there has been a
controversy if the kinematic or rather the normal stress boundary condition
should be used to move the free surface. This issue was resolved by Silliman and
Scriven who state that for capillary numbers below unity, the normal stress
iteration converges well while a kinematic iteration eventually
fails.\cite{SaiScr81} In addition, when the kinematic boundary condition is used
for updating the free surface, the balance of normal stress that carries the
effects of surface tension is not strictly imposed. It is used when implementing
the weak form of the Navier--Stokes equations: In this context an integration by
parts yields an integral of the normal stress over the free surface, which is
then replaced by the corresponding surface-integral of the tension forces.
Similar techniques are commonly used for problems with outflow boundary
conditions or for Poisson's equation with Neumann boundary conditions. The
correctness of the technique has been justified for the outflow problem by
Renardy.\cite{Renardy97} However, it is not evident if it also works in the case
where the surface-tension terms dominate the whole problem. The question remains
open, in which sense the boundary condition is satisfied. Therefore, we found it
necessary in our examples to visualize the terms involved in the free-surface
boundary condition, thus proving that they are correctly balanced.

An important result of a variational description of the tension terms is an
improvement of the Newton algorithm controlling possible mesh distortions at the
free surface. Many algorithms implementing the weak form of the capillary
boundary condition encounter intrinsic instabilities of the boundary mesh when
significant changes of the free surface take place. For the program
\emph{surface evolver} \cite{Brakke92} this manifests itself in shrinking and
growing surface facets. Similar effects have been observed by
Brinkmann\cite{Brinkmann02} and B\"ansch.\cite{Bansch98} Our formulation of the
capillary free surface is such that the free surface smoothly changes to the
behavior of a rubber band when the boundary mesh becomes distorted. This leads
to an automatic regularization of the mesh without the need of explicit
remeshing or smoothing.

In Section~\secref{sec:problem} the mathematical formulation of the problem is
presented in terms of differential equations, together with the boundary
conditions and the relevant parameter regime. In Sec.~\secref{sec:continuous} we
then re-establish the bulk equations and their boundary conditions by
variational techniques. For the free-surface we introduce a
differential-geometric notation that allows us to write the boundary condition
in a weak form. Up to this point a continuous description is used.
Section~\secref{sec:discretization} introduces the discretization of the problem
by the computational mesh. The formulation of the tension forces as the
concurrent minimization of the free-surface area of single finite elements is a
necessary requirement for the mentioned automatic regularization mechanism.
Sec.~\secref{sec:summary} provides a short summary of the whole algorithm. In
Sec.~\secref{sec:numerics} we present examples that show the accuracy of the
algorithm and two further examples for different values of the capillary and
Bond numbers. Mathematical and algorithmic details are deferred into the
Appendices~\secref{sec:static}--\secref{sec:slipconstraints}.

\section{Statement of the problem} 
\label{sec:problem}%

Throughout the paper, we shall write all equations in tensor notation for
arbitrary curvilinear coordinate systems. This will considerably simplify the
differential geometric notation in the following sections. For the formulation
of the full Navier--Stokes equations in curvilinear coordinates we refer to
Aris.\cite{Aris89} A repeated index that occurs in co- and contravariant
positions is summed over, indices that are preceded by a comma denote covariant
derivatives, and $g_{ij}$~is the metric tensor of the underlying coordinate
system.

\subsection{The basic equations}

We study incompressible and stationary flows, that are characterized by a small
Reynolds number $\Reynolds = \rho\bar x\bar v/\eta$. Here, $\rho$~is the density
of the fluid, $\viscosity$~its viscosity, and $\bar x$ and $\bar v$ denote
typical magnitudes of length and velocity. Under these conditions the pressure
field~$p$ and the velocity field with components~$v^i$ satisfy the Stokes
equations~\cite{LanLif63}
\begin{gather}
  \label{eq:divfree}
  v^i_{,i} = 0, \\
  \label{eq:stokes}
  0 = \stress^{ij}_{,j} + f^i \quad
  \text{with}\quad
  \stress_{ij} = -p g_{ij} + \viscosity (v_{i,j} + v_{j,i}),
\end{gather}
where $\stress_{ij}$~is the fluidic stress tensor, $f^i$~an external body force
causing non-trivial streaming and pressure patterns within a domain~$V$. It can
be split into its conservative part~$f^i_\text{c}$ which can be displayed as the
gradient of a potential and its non-conservative part~$f^i_\text{nc}$ with
vanishing divergence. The domain~$V$ may be bounded by rigid walls and by free
surfaces, as e.g.\ a droplet sitting on a substrate. Equations
\eqnref{eq:divfree}~and \eqnref{eq:stokes} then are subject to boundary
conditions at different parts of the boundary~$\partial V$: First, the flow
has to meet the \emph{kinematic} boundary condition, requiring that the normal
projection of a stationary velocity field vanishes at the boundary, i.e.,
\begin{equation}
  \label{eq:kinematic-bc}
  v_i N^i = 0.
\end{equation}
At immobile sticky walls we use the \emph{no-slip} boundary condition, according
to which the velocity vanishes also in the tangential directions of the
boundary, implying
\begin{equation}
  \label{eq:noslip-bc}
  v_i T^i_{\alpha} = 0 \quad\text{at the walls.}
\end{equation}
Here, $T^i_{\alpha}$~denotes the $i$th component of the tangential
vector~$\bfT_{\alpha}$ ($\alpha=1,2$ for a two-dimensional surface). The
remaining boundary is a free surface that dynamically adjusts its position such
that the stress balance holds,
\begin{equation}
  \label{eq:free-bc}
  \stress^{ij} N_j = \tension\curv N^i \quad\text{on free surfaces,}
\end{equation}
with the surface tension~$\tension$ and the curvature~$\curv$. Note that we have
omitted the gradient of the surface tension and thus exclude Marangoni effects.
This simplifies the following calculations but does not present a principal
restriction of our description.

Equations~\eqnref{eq:divfree}--\eqnref{eq:free-bc} have been simplified by
assuming that the fluid in domain~$V$ is surrounded by a medium of much smaller
viscosity, which is the case e.g.\ for a water--air interface at room
temperature. Therefore, the surrounding's viscous stress contribution does not
show up in the balance Eq.~\eqnref{eq:free-bc}. We further assume that the
ambient pressure $p_0$~is homogeneous. Since the pressure is determined by the
Stokes equations only up to a constant, we can split it into a part~$p_1$ with
vanishing average and use the ambient pressure~$p_0$ as an offset parameter
which enters only in the normal stress balance~\eqnref{eq:free-bc},
\begin{equation}
  \label{eq:p-split}
  p(\bfx) = p_0 + p_1(\bfx) \quad\text{with}\quad
  \int_V p_1\,dV = 0.
\end{equation}
%

\subsection{The parameter regime}

By transforming both, the bulk equation~\eqnref{eq:stokes} and the free boundary
condition~\eqnref{eq:free-bc} into dimensionless form employing
viscosity-scaling, one observes that the system may be characterized by two
relevant ratios of forces, given by the dimensionless numbers
\begin{equation}
  \Bond = \frac{\bar{f}_\text{c}\bar x^2}{\tension}\quad\text{and}\quad
  \Capillary = \frac{\viscosity\bar{v}}{\tension}.
\end{equation}
Here, $\bar{x}, \bar{v}$ and~$\bar{f}_\text{c}$ denote typical magnitudes of
length, velocity and the conservative part of the force density, respectively.
$\Bond$ is a generalization of the \emph{Bond number} which is usually defined
in terms of gravitational forces only. The \emph{capillary number}~$\Capillary$
measures the viscous contribution to the surface deformation. In a system with
static boundaries and vanishing \emph{Reynolds number} we can express the
velocity scale by the typical magnitude~$\bar{f}_\text{nc}$ of the
non-conservative part of the driving force, namely $\bar{v} =
\bar{x}^2\bar{f}_\text{nc}/\viscosity$. This yields an alternative definition of
the capillary number similar to that of the Bond number,
\begin{equation}
  \Capillary = \frac{\bar{f}_\text{nc}\bar x^2}{\tension}.
\end{equation}
These two numbers reflect the very different effects of the conservative and the
non-conservative parts of the driving. In this sense, $\Capillary$~provides also
a measure for the spatial changes of the velocity field. For small~$\Capillary$
the flow is slow and changes smoothly, whereas for large~$\Capillary$ it may
exhibit drastic gradients.

We propose our numerical scheme for the parameter regime where both,
$\Capillary$, and $\Bond$ are of order unity or less. Thus, pressure gradients
and viscous forces can deform the free surface significantly. The surface
tension is large enough, however, in order to keep the whole fluid domain
together, a pinch-off cannot occur. The viscosity renders the velocity field
smooth over the whole fluid domain and prevents the existence of boundary
layers. Because we consider only stationary flows in stationary domains
according to Eqs.~\eqnref{eq:stokes}~and \eqnref{eq:kinematic-bc}, and because
we have set the substrate's velocity to zero in Eq.~\eqnref{eq:noslip-bc}, we
always obtain \emph{pinned} contact-lines. A rolling or slipping droplet would
raise additional challenges regarding the stress near the contact-line that are
beyond the scope of this paper.

\section{Continuous description of the problem}
\label{sec:continuous}

In order to clarify the numerical treatment of the free-surface boundary
condition we first explore the physical origins of the balanced forces. We will
then express each of them by the first variation of a functional. For Newton's
method it will be necessary to calculate also the second variation.

\subsection{Physical aspects of the free-surface boundary condition: first variations}
\label{sec:physical}%

The surface tension term~$\tension\kappa N_i$ in the boundary
condition~\eqnref{eq:free-bc} arises from the fact that an extended interface
between two different phases ``costs'' free energy.\cite{LanLif63} To find the
optimal configuration the surface is continually probing positions in its
vicinity in order to minimize its free energy. For the case of an applied
conservative force~$f_i = -\Phi_{,i}$ the system is static ($v^i = 0$), and the
free-surface boundary condition is equivalent to a minimization of a free energy
expression. This calculation is performed in Appendix~\secref{sec:static}.

\begin{figure}%
  \centerline{\includegraphics{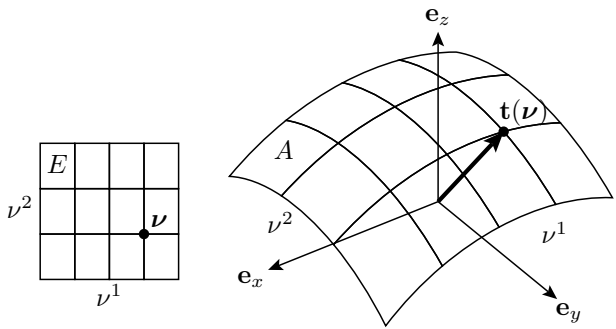}}%
  \caption{A sketch of the coordinate system on a two-dimensional surface~$A$,
  embedded into the three-dimensional space. The surface coordinates~$\bfnu$ are
  mapped from the reference domain~$E$ (left) onto the surface~$A$ (right) via
  the parameterization vector~$\bft(\bfnu)$.}%
  \label{fig:coordinates}%
\end{figure}%
Due to its thermodynamic origin, the surface tension term results from a first
variation of a functional. This carries over also to the dynamic case ($v^i \neq
0$) in which the boundary condition~\eqnref{eq:free-bc} must hold at any instant
of time. The contribution of the free surface~$A$ to the system's free energy is
given by the integral of the surface tension~$\tension$ over~$A$, where
$dA$~denotes the infinitesimal surface area,
\begin{equation}
  \label{eq:func-tens}
  F = \int_A \tension\:dA.
\end{equation}
Any smooth surface in a $\spacedim$-dimensional space may be parameterized by
$\spacedim{-}1$ surface coordinates~$\nu^\alpha$
($\alpha=1,\ldots,\spacedim{-}1$) which determine the
coordinates~$t^i(\nu^\alpha)$ of points in $\spacedim$-dimensional space on the
surface. Both, surface and space coordinates are illustrated in
Fig.~\figref{fig:coordinates}. In our numerical studies we restrict
ourselves to $\spacedim=2$. The general framework, however, remains valid also
for~$\spacedim=3$. The surface coordinates~$\nu^\alpha$ are taken from the
parameter set~$E\subset\Real^{\spacedim{-}1}$. With $\bfnu$~running through~$E$
the whole free surface~$A$ is covered,
\begin{gather}
  \label{eq:parameterization}
  t^i\colon E\to \Real\colon \bfnu\mapsto t^i(\bfnu), \\
  A = \{\bfe_{(i)} t^i(\bfnu) \mid \bfnu\in E \}.
\end{gather}
Here, $\bfe_{(i)}$~is the $i$th base vector in space. Throughout the paper we
will always use Greek symbols for surface indices and Latin ones for space
indices. The connection between surface and space coordinates is conveniently
described by the surface-derivatives of the parameterization
functions (cf. Aris,\cite{Aris89} p.~215), i.~e.,
\begin{equation}
  t^i_{,\alpha}(\nu^\beta) = \frac{\partial t^i}{\partial \nu^\alpha} (\nu^\beta).
\end{equation}
Understood as a contravariant $\spacedim$-dimensional space-vector,
$t^i_{,\alpha}$ represent the components of the $\alpha$th~tangent
vector~$\bfT_\alpha$ to the surface. At the same time, $t^i_{,\alpha}$~is a
covariant surface-vector. We can now construct the components of the surface's
metric tensor~$a_{ij}$ as the scalar products of these tangent vectors, namely
\begin{equation}
  a_{\alpha\beta} = g_{ij}\, t^i_{,\alpha}\, t^j_{,\beta} \quad\text{and}\quad
  a = \det(a_{\alpha\beta}).
\end{equation}
The metric tensor~$a_{\alpha\beta}$, its determinant~$a$ and its
inverse~$a^{\alpha\beta}$ are nonlinear functions of the tangential vector
components~$t^i_{,\alpha}$, see in Appendix~\secref{sec:diffgeom-var}. The
normal vector the normalized cross-product of two tangent vectors,
\begin{equation}
  \label{eq:normal}
  N_i = \frac{1}{2}\varepsilon_{ijk}\,\varepsilon^{\alpha\beta}\,
        t^j_{,\alpha}\, t^k_{,\beta},
\end{equation}
and the curvature~$\kappa$ is given as the trace of the tensor~$b_{\alpha\beta}$
of the second fundamental form of the surface,
\begin{gather}
  \label{eq:curvature}
  \kappa = a^{\alpha\beta} b_{\alpha\beta}\quad\text{with}\\
  \label{eq:secondfundf}
  b_{\alpha\beta} = t^i_{,\alpha\beta} N_i.
\end{gather}
Using the parameterization~\eqnref{eq:parameterization} of the free surface, we
demonstrate in Appendix~\secref{sec:diffgeom-var} that the change of the free
energy contribution~$F$ with respect to a variation of the surface vectors~$t^i$
is given~by
\begin{equation}
  \label{eq:var-t}
  \variat F [\variat\bft] = \int_A \tension\,t^j_{,\beta}\, g_{ij}\,
  a^{\alpha\beta}\, \variat t^i_{,\alpha} \:dA.
\end{equation}
By an integration by parts this expression can be cast into a form containing the
curvature term of the free-surface boundary condition~\eqnref{eq:free-bc},\
i.e.,
\begin{equation}
  \label{eq:var-t-partint}
  \variat F [\variat\bft]
    = - \int_A \tension\curv N_i \variat t^i \,dA
\end{equation}
(see Appendix~\secref{sec:curvature} also for the case of varying surface
tension). In this way, the curvature term~$\curv$ in
Eq.~\eqnref{eq:var-t-partint} that contains second spatial derivatives is
replaced by a product of two terms, each containing a first derivative in
Eq.~\eqnref{eq:var-t}. Especially for numerical applications it is much more
favorable to work only with first derivatives. This trick has been used in the
literature in different
contexts.\cite{ZinRotDav97,Brakke92,Dziuk91,DecSie00,Bansch98} Seen from a
physical perspective, version~\eqnref{eq:var-t} of the equation is the more
natural one. Here, one directly deduces that forces pulling along the tangential
direction attempt to minimize the facet area of a mesh's boundary. On the basis
of single finite elements this perspective will be used below for stabilizing
the computational mesh.

The left-hand side of Eq.~\eqnref{eq:free-bc}, $\sigma_{ij} N^j$, is the normal
fluidic stress at the boundary. We now recapitulate how this term can be
understood as the result of a variational principle. In a stationary system with
rigid immobile boundaries there is a balance between the power output due to
\emph{viscous dissipation} and the power input due to \emph{external driving}.
The total power output of the fluid, which we will denote with
\begin{equation}
  P
  = \int_V \calP \,dV
  = \int_V (\sigma^{ij} v_{i,j} - f^i v_i) \,dV,
\end{equation}
vanishes. Additionally, for a prescribed domain~$V$, the Stokes equations yield
those velocity and pressure fields that render the local power extremal (see
Finlayson,\cite{Finlayson72} p.~271). Vice versa, from the vanishing first
variation of~$P$,
\begin{align}
  \label{eq:var-p}
  \variat P[\variat p]
     &= \int_V \frac{\formal\calP}{\formal p} \variat p \,dV
      = - \int_V v^i_{,i} \variat p \,dV\\
  \variat P[\variat \bfv]
     &=\int_V \Bigl(\frac{\formal\calP}{\formal v_i} \variat v_i
                  + \frac{\formal\calP}{\formal v_{i,j}} \variat
                  v_{i,j}\Bigr)\,dV \\
     &\mkern-30mu= \int_V (-f^i\variat v_i + \sigma^{ij} \variat v_{i,j}) \,dV \\
     \label{eq:var-v}
     &\mkern-30mu= - \int_V (f^i + \sigma^{ij}_{,j}) \variat v_i \,dV
       + \oint_{\partial V} \sigma^{ij} N_j \variat v_i \,dA,
\end{align}
the Stokes equations follow by setting the bulk contributions to zero. The
boundary integral in the last row of Eq.~\eqnref{eq:var-v} provides the fluidic
stress in the free-surface boundary condition.

At this point we see that the two terms in the stress balance
Eq.~\eqnref{eq:free-bc} have different physical origins. The surface tension is
of thermodynamic (or rather of ``thermostatic'') nature while the fluidic stress
stems from dynamic considerations. The first results from minimizing a free
energy, while the second stems from minimizing a power. Formally, this is
expressed by the different variations $\variat v^i$~and $\variat t^i$ in the
expressions $\sigma_{ij}N^j\variat v^i$ in Eq.~\eqnref{eq:var-v} and
$\curv\tension N_i \variat t^i$ in Eq.~\eqnref{eq:var-t-partint}. Already for
dimensionality reasons they cannot be equal, neither can the functionals $F$~and
$P$ be directly combined into one single variational principle.

From an algorithmic point of view one has to make a choice here: to approximate
the free-surface boundary condition using either the~$\variat v^i$ or
the~$\variat t^i$ as test-functions. In our Galerkin implementation of the
problem we will use the ansatz functions as test functions. Therefore, in order
to acquire a consistent numerical algorithm we have to approximate both,
velocity and the geometry parameterization with finite elements of the very same
order. This is the first central statement of the present work.

It was stated by B\"ansch\cite{Bansch98} (p.~42, cf.~also citations 49~and 50
therein) that a second-order approximation of the surface parameterization
yields a ``good discrete curvature'', whereas a first-order one does not. The
same can be seen below in Fig.~\figref{fig:circle}. We are now able to
substantiate his numerical observation with the underlying physical mechanism.
The argument is similar to that for the celebrated
Ladyzhenskaya--Babuska--Brezzi requirement that velocity gradients have to be
approximated by the same order as the pressure. From a physical perspective this
is not astonishing, because both are components of the same stress tensor.

\subsection{Splitting the problem into two numerical systems}
\label{sec:splitting}%

For free boundaries a twofold problem must be solved: (i) The unknown fluid
domain~$V$ is to be determined and (ii) the Stokes Eqs.\ \eqnref{eq:divfree}~and
\eqnref{eq:stokes} are to be solved within~$V$, using the boundary conditions
\eqnref{eq:kinematic-bc}--\eqnref{eq:free-bc}. The latter themselves depend on
the shape of~$V$ via the normal vector at the boundary. Both parts of this
problem cannot be processed independently.

In principle, there exist two options to deal with this combined problem. A
first one is to implement a single numerical system for both, the flow variables
$p$~and $v^i$ together with the geometry variables~$t^i$. We will not follow
this direction but rather consecutively solve two smaller systems, one for the
flow variables, depending on the current domain~$V$, and a second one for the
parameterization of the boundary. We have chosen this approach because the
problem is linear in the flow variables and highly nonlinear in the geometry
variables~$t^i$. The nonlinearity is due to the appearance of the inverse
surface metric~$a^{\alpha\beta}$ in Eq.~\eqnref{eq:var-t}. Thus, solving the
Stokes equations in the \emph{fluidic system}, how we will call it, will be a
standard problem, while the nonlinear search for the correct boundary shape will
be done in the \emph{geometry system}. Both systems are solved consecutively:
\begin{enumerate}
  \renewcommand{\theenumi}{\arabic{enumi}}%
  \item Choose an initial domain~$V$.
  \item Until convergence repeat the following steps:
  \begin{enumerate}
    \makeatletter
    \renewcommand{\theenumii}{{\normalfont\itshape\@alph\c@enumii\/}}%
    \makeatother
    \item Solve the fluidic system within the domain~$V$.
    \item Solve the geometry system using fixed values for the pressure and
    velocity variables. This results in an updated domain~$V$.
  \end{enumerate}
\end{enumerate}

In three-dimensional space the equations
\eqnref{eq:kinematic-bc}--\eqnref{eq:free-bc} pose four boundary conditions.
They are one too many for the linear fluidic system to be fully determined. One
boundary condition is thus used for updating the parameterization of the free
surface.\cite{CuvSch90} The main challenge is the proper assignment of specific
boundary conditions to the two systems in order to make them solvable, uniquely
determined, and robust. It is clear that the no-slip boundary
condition~\eqnref{eq:noslip-bc} at sticky walls applies only to the fluidic
system. The free-surface boundary condition yet needs further consideration.

Here, again, a physical argument helps to choose the proper boundary condition.
It is either the stress by the fluid or its velocity that is moving the free
surface. Accordingly, either the normal stress balance~\eqnref{eq:free-bc} or
the kinematic boundary condition~\eqnref{eq:kinematic-bc} can be used by the
geometric system to update the surface (see the discussion by Saito~\&
Scriven\cite{SaiScr81} and our remarks in the introduction). We choose our
approach according to the following principle: The fluidic system should be well
defined as a stationary system even if the boundary is fixed and is not part of
the problem. If then the kinematic boundary condition were not imposed on the
stationary flow, the velocity field would pass through the free surface which is
also stationary. This excludes surface-updates by the kinematic boundary
condition.

Until the correct boundary shape has been found, it is, in principle, possible
that the surrounding flow forces the free surface into an arbitrary direction.
By its very nature, however, the tension force stays always normal on the free
surface. Only normal forces can be compensated by a free surface. As a necessary
condition the tangential projection of the normal stress has to
vanish.\cite{ill-posed} Whenever tangential components emerge during the run of
an algorithm, the result will be a numerical artefact. In the proposed scheme
with two separated systems it is the fluidic system which must ensure the
tangential components of the free boundary condition, i.e.,
\begin{equation}
  \label{eq:free-bc-t}
  (v_{i,j} + v_{j,i}) N^i t^j_{,\alpha} = 0
  \quad\text{for all $\alpha$}.
\end{equation}
Here, the surface tension $\tension$~has been set constant along the surface.
For the velocity variables this constitutes a perfect slip boundary condition,
which is similar to a Neumann boundary condition. We thus find the fluidic
system to be fully determined and physically well defined even for fixed
boundaries by the conditions~\eqnref{eq:kinematic-bc}, \eqnref{eq:noslip-bc}~and
\eqnref{eq:free-bc-t}.

The geometry system is then responsible for the remaining normal component of
the stress balance~\eqnref{eq:free-bc},
\begin{equation}
  \label{eq:free-bc-n}
  -p + \viscosity (v_{i,j} + v_{j,i}) N^i N^j = \tension\curv,
\end{equation}
which is used as the update equation for the boundary. The free surface moves
if Eq.~\eqnref{eq:free-bc-n} does not hold for a given trial boundary.

\subsection{Second variation with respect to the surface parameterization}
\label{sec:variation}%

In a first implementation we used a direct and explicit update algorithm moving
the boundary into normal direction with a step-width that is determined by a
parameter~$\tau$ and the residual of Eq.~\eqnref{eq:free-bc-n}. The
discretization of this update can be found below in
Eq.~\eqnref{eq:direct-update}. Depending on the value of~$\tau$ this method
exhibited strong instabilities as demonstrated below in
Fig.~\figref{fig:explicit}. Although advanced techniques for determining an apt
value for~$\tau$ seem to exist (cf.\ the program \emph{surface evolver} by
Brakke\cite{Brakke92}), we prefer a Newton--Raphson iterative method. This has
the advantage of a faster convergence and a less strong dependence on~$\tau$. A
minor disadvantage is that it requires an additional variation of the surface
free energy for the assembly of the geometry system. Using the same calculus as
in Appendix~\secref{sec:diffgeom-var} we find the second variation of the free
energy contribution~$F$ of a one-dimensional free surface,
\begin{align}
  \variat^2 F&[\variat\bft,\variat\bft]
   = \variat\left(\int_A \tension\: g_{ij} a^{\alpha\beta} t^j_{,\beta}
     \variat t^i_{,\alpha} \,dA \right)\\
  \label{eq:var-tt}
  &\begin{aligned}[t]
   &= \int_A \tension\: \variat t^i_{,\alpha} g_{ij} a^{\alpha\beta}
            \variat t^j_{,\beta} \, dA \\
   & -\int_A \tension\: (\variat t^i_{,\alpha} \, g_{ik} \, a^{\alpha\psi}
            \, t^k_{,\psi}) \: (\variat t^j_{,\beta}  \, g_{jl} \, a^{\beta\phi}
            \, t^l_{,\phi}) \, dA.
  \end{aligned}
\end{align}
For a two-dimensional surface the corresponding variation contains two
additional terms that are not given here for brevity. The last integral in
Eq.~\eqnref{eq:var-tt} turns out to cause numeric instabilities in Newton's
method. This is a rather surprising fact, because the calculation that led to
Eq.~\eqnref{eq:var-tt} consists of two straightforward variations. If the last
integral in Eq.~\eqnref{eq:var-tt} is omitted the algorithm becomes stable and
accurate (cf.~the tests in Sec.~\secref{sec:accuracy}).

It is not only the free energy contribution~$F$ that depends on the shape of the
surface. Also, the flow velocity, and by this, the viscous stress and the
pressure depend on the shape. The formulation of the Newton method
requires also the change of the fluidic stress integral due to changes of the
free boundary,
\begin{multline}
  \label{eq:variatstress}
  \variat\left(\int_A \stress_{ij}N^j\variat t^i\,dA\right)[\variat\bft]
  = \int_A \variat t^i \sigma_{ij} \variat N^j[\variat\bft] \,dA \\
  + \int_E \variat t^i \sigma_{ij} N^j \variat\sqrt{a}[\variat\bft] \,d\bfnu
  + \int_A \variat t^i \variat \sigma_{ij}[\variat\bft] \,N^j \,dA.
\end{multline}
The first two integrals on the right-hand side contain the changes of the normal
vector~\eqnref{eq:normal} and the infinitesimal surface area~$dA =
\sqrt{a}d\bfnu$ due to changes of the boundary's shape. Both can be calculated
along the lines of Appendix~\secref{sec:diffgeom-var}. The third integral
expresses the change of the fluidic stress~$\stress_{ij}$ at the boundary due to
changes of its position. The shape changes are communicated to the flow and
pressure fields via the boundary conditions \eqnref{eq:kinematic-bc}~and
\eqnref{eq:free-bc-t} of the Stokes equations. Unfortunately, this very indirect
response of the stress tensor on the changes of shape cannot be expressed
exactly. We therefore have to assume that this term can expressed by derivatives
of the stress tensor, i.e.,
\begin{equation}
  \label{eq:stressvariat}
  \variat\sigma_{ij}[\variat\bft] \approx \sigma_{ij,k} \variat t^k.
\end{equation}
This means that the fluidic and the geometry system decouple to the extent that
the stress tensor in the vicinity of the boundary is not affected by small
boundary changes. We note that this is not a consequence of splitting the
problem into two separate systems, but a general problem that equally applies to
the combined approach. Altogether, the right-hand side of
Eq.~\eqnref{eq:variatstress} becomes approximately
\begin{multline}
    \int_A
    \Bigl\{(\variat t^i \sigma_{ij} N^j)
           (\variat t^k_{,\alpha} g_{kl}a^{\alpha\beta} t^l_{,\beta}) \\
         - (\variat t^i \sigma_{ij} t^j_{,\alpha}) a^{\alpha\beta}
           (\variat t^k_{,\beta} N_k)
         + \variat t^i \sigma_{ij,k} N^j \variat t^k
     \Bigr\} \, dA.
\end{multline}
%


\section{Discretization of the problem}
\label{sec:discretization}%

We implemented the above equations by means of a Galerkin approximation scheme
which is known to work well for minimization problems. As variables we
introduced the velocity components $u$~and $v$ in $x$- and $y$-direction,
respectively, the pressure~$p$, and additional variables $r$~and $s$ for the
coordinates of the boundary parameterization vector~$\bft$. The continuous
fields are discretized using ansatz functions, weighted with the corresponding
degrees-of-freedom~(DoF),
\begin{align}
  \label{eq:ansatzphi}
  u(\bfx) &= \sum_{d} u_d \phi_d(\bfx),\qquad
  v(\bfx)  = \sum_{d} v_d \phi_d(\bfx),\\
  p(\bfx) &= \sum_{d} p_d \psi_d(\bfx),\\
  r(\bfx) &= \sum_{d} r_d \chi_d(\bfx),\qquad
  s(\bfx)  = \sum_{d} s_d \chi_d(\bfx),
\end{align}
where the sum runs over all DoFs. The fluid velocity components~$u,v$ are
approximated by the second-order finite elements (FEs)~$\phi$ and the pressure
variable~$p$ by first-order FEs~$\psi$. For the position variables~$r,s$ we have
predominantly used second-order FEs, but for accuracy and other testing reasons
we also tried first-order FEs. We denote the position FEs with~$\chi$. All FEs
are of the Lagrange family,\cite{DevFisMun02} having ansatz functions that
are~$1$ at exactly one node of the mesh and $0$~at all others. The DoFs are then
equal to the function values at the nodes. This property is most convenient for
the position variables~$(r_d, s_d)$ that coincide with the coordinates of the
node~$d$.

\subsection{The fluidic system}
The fluidic system is implemented in a standard way. Equation~\eqnref{eq:stokes}
is tested with the second-order FEs~$\phi$, while the continuity
Eq.~\eqnref{eq:divfree} is tested with the first-order FEs~$\psi$. We have
implemented the following linear equation for the DoFs, which are collected to
vectors $\vecu,\vecv,\vecp$ with components $u_d,v_d,p_d$ respectively,
\begin{equation}
  \label{eq:fluidicmatrix}
  \left(\begin{array}{ccc}
    K_{uu} & 0      & K_{up} \\
    0      & K_{vv} & K_{vp} \\
    K_{pu} & K_{pv} & 0
  \end{array}\right)
  \left(\begin{array}{c}
    \vecu \\ \vecv \\ \vecp
  \end{array}\right)
  =
  \left(\begin{array}{c}
    L_u \\ L_v \\ 0
  \end{array}\right)
\end{equation}
with the entry matrices $K$ and entry vectors $L$ given by
\begin{align}
  \label{eq:fluidicintegrals-begin}
  [K_{uu}]_{de} &=  \viscosity\!\!\int\limits_V \bfnabla\phi_d\scalarmult\bfnabla\phi_e\,dV
                   - \viscosity\!\!\oint\limits _{\partial V} \phi_d \bfN\scalarmult\bfnabla\phi_e\,dA, \\
  [K_{vv}]_{de} &=  \viscosity\!\!\int\limits_V \bfnabla\phi_d\scalarmult\bfnabla\phi_e\,dV
                   - \viscosity\!\!\oint\limits_{\partial V} \phi_d \bfN\scalarmult\bfnabla\phi_e\,dA, \\
  [K_{up}]_{de} &= -\int_V (\partial_x\phi_d)\psi_e\,dV
                   + \oint_{\partial V} \phi_d\psi_e N_x\,dA, \\
  [K_{vp}]_{de} &= -\int_V (\partial_y\phi_d)\psi_e\,dV
                   + \oint_{\partial V} \phi_d\psi_e N_y\,dA, \\
  [K_{pu}]_{de} &= -\int_V \psi_d\partial_x\phi_e \,dV,\\
  [K_{pv}]_{de} &= -\int_V \psi_d\partial_y\phi_e \,dV,\\
  [L_u]_d       &=  \int_V \phi_d f_x\,dV, \\
  [L_v]_d       &=  \int_V \phi_d f_y\,dV.
  \label{eq:fluidicintegrals-end}
\end{align}
All integrals are assembled in a loop over the elements and the sides of the
mesh, using a fifth-order Gaussian quadrature rule. The fluidic system could
likewise implement the stationary Navier--Stokes equations with a small Reynolds
number; we have chosen the Stokes equation for simplicity reasons here.

The boundary conditions are imposed by a constraints technique for the matrix and
for the right-hand side in Eq.~\eqnref{eq:fluidicmatrix}. A constrained
DoF~$u_d$ is expressed by an inhomogeneity plus a weighted sum of other DoFs,
\begin{equation}
  \label{eq:constraint}
  u_d = w_d + \sum_{e\neq d} w_{de} u_e,
\end{equation}
which represent the boundary condition in question. The DoF~$u_d$ is then
completely eliminated from the linear system~\eqnref{eq:fluidicmatrix}. By such
constraint equations we implemented weak formulations of the kinematic boundary
condition~\eqnref{eq:kinematic-bc}, i.~e.,
\begin{equation}
  \label{eq:kinematic-bc-w}
  0 = \sum_e (u_e N_x + v_e N_y) \int_{\partial V} \phi_d \phi_e \, dA,
\end{equation}
of the no-slip condition at the walls
\begin{equation}
  \label{eq:noslip-bc-w}
  0 = \sum_e (u_e T_x + v_e T_y) \int_{\partial V} \phi_d \phi_e \, dA,
\end{equation}
and of the tangential projection of the free-surface boundary
condition~\eqnref{eq:free-bc-t},
\begin{multline}
  \label{eq:free-bc-t-w}
  \arraycolsep=0pt
  0 = \sum_e
      \left(\begin{array}{c} u_e\\v_e \end{array}\right)
      \scalarmult
      \left(\begin{array}{c} 2T_xN_x \\ T_xN_y{+}T_yN_x \end{array}
      \begin{array}{c}T_xN_y{+}T_yN_x \\ 2T_yN_y \end{array}\right)\,\times\\
      \times\,\int_{\partial V} \phi_d
      \left(\begin{array}{c} \partial_x\phi_e\\
      \partial_y\phi_e\end{array}\right) \, dA.
\end{multline}
The constraint equations differ only in the values of~$w_{de}$. The
inhomogeneity~$w_d$ is zero in all three equations. Non-zero inhomogeneities
would result, if also a surface-gradient term of the tension were taken into
account in Eq.~\eqnref{eq:free-bc-t}, or if the rigid walls performed a
tangential movement.

For the boundary condition~\eqnref{eq:free-bc-t-w}, which is equivalent to an
ideal slip condition, it is known that an improper choice of the normal
direction can cause spurious contributions to the velocity field (see
Behr,\cite{Behr04} Walkley~\etal,\cite{WalGasJimKelSum04} and our remarks stated
in the introduction). In the presence of conservative forces only we did not
find such spurious flows in our results.

The fact that the formulation of the free boundary condition in terms of the
DoF-constraints~\eqnref{eq:free-bc-t-w} cross-links all DoFs residing at
boundary nodes, presents a serious problem. Each of the DoFs is in principle
linked to all its neighbors on the boundary. This leads to a nearly filled
system-matrix which is unfavorable regarding memory capacity and computing
time. We found that an iterative method can overcome this problem. Instead of
cross-linking a boundary DoF with all its neighbors, for some of them we take
their old values, as is detailed in Appendix~\secref{sec:slipconstraints}. After
some iterations the full boundary condition~\eqnref{eq:free-bc-t-w} is
established. The drawback of this scheme is that the constraint equations have
to be re-assembled after every solution step of the fluidic system.

\subsection{The geometry system}
The geometry system employs a Newton method to perform the nonlinear search for
the correct boundary position. This scheme corresponds to a minimization of the
free energy~$F$, while taking the fluidic stress into account. The boundary update
equation can be written in a discretized form as
\begin{equation}
  \begin{aligned}
    0 &= [L_r]_d(\vecr,\vecs,\vecu,\vecv,\vecp)
      := \frac{\partial F}{\partial x_d} + \int_A \chi_d \sigma^{xj}N_j \,dA, \\
    0 &= [L_s]_d(\vecr,\vecs,\vecu,\vecv,\vecp)
      := \frac{\partial F}{\partial y_d} + \int_A \chi_d \sigma^{yj}N_j \,dA,
  \end{aligned}
\end{equation}
where $d$~runs over the DoFs for each geometry variable, and $L$~and $F$ are
understood as functions of the arrays $\vecr,\vecs$, etc.~containing the~DoFs. For the
Newton--Raphson method the geometry system repeatedly has to solve the linear
system of equations~\cite{BroSemMus95}
\begin{multline}
  \label{eq:geometricmatrix}
  \left(\begin{array}{cc}
    \partial [L_r]_d / \partial r_e &
    \partial [L_r]_d / \partial s_e \\
    \partial [L_s]_d / \partial r_e &
    \partial [L_s]_d / \partial s_e
  \end{array}\right)^\text{(old)}
  \left(\begin{array}{c}
    \vecr_e^\text{\,(new)} - \vecr_e^\text{\,(old)} \\
    \vecs_e^\text{\,(new)} - \vecs_e^\text{\,(old)}
  \end{array}\right) \\
  = - \tau
  \left(\begin{array}{c}
    [L_r]_d \\{}
    [L_s]_d
  \end{array}\right)^\text{(old)}
\end{multline}
where $\tau\in[0,1]$~is a step-size parameter. In all applications we have used
values of~$\tau$ between $0.1$~and $1.0$.

The search for the correct boundary shape is strongly nonlinear in the position
variables. In order to remove the main nonlinearities, which are caused by the
surface metric expressions $\sqrt{a}$~and $a^{\alpha\beta}$, the nodes of the
elements are moved to their corresponding coordinates~$(r_d, s_d)$ after each
step of the geometry system. Then, all integrals can be performed directly on
the elements' edges. Also the normal vector can be taken from the elements'
sides. In the previous section we used a convenient variational notation to
express the change of the free energy contribution~$F$ by changes of the
boundary parameterization. Essentially the same equations are obtained by
differentiating the discrete version of~$F$ with respect to the DoFs which are
the nodal degrees of freedom of the corresponding variables. The only difference
is that the variation $\variat t^i$ in the continuous formulation must be
replaced by the vectorial test-function $\chi_d \bfe_i$, and the variation
$\variat t^i_{,\alpha}$ by its tangential derivative
$\bfT_\alpha\scalarmult\bfnabla\chi_d \bfe_i$.

\subsection{Controlling the tangential displacements of boundary nodes}
\label{sec:rubber}

For a given discretization we must not only find the correct boundary
shape, but its discretization should also remain well-proportionate. Very long
and very short element sides cause badly conditioned matrices and make the whole
algorithm unstable. Several algorithms implementing the weak form of the
free-surface boundary condition encounter these intrinsic instabilities of the
boundary mesh. For the program \emph{surface evolver} this manifests itself in
shrinking and growing surface facets. It is therefore recommended to monitor the
mesh quality and remove too small or split too large elements.\cite{Brakke92}
Similar effects were reported by Brinkmann.\cite{Brinkmann02}

In Sec.~\secref{sec:splitting} the assignment of the boundary conditions to the
fluidic and the geometry systems was described. There, we found that the
presence of incompatible forces may easily destroy a free surface which
essentially attempts to minimize the lengths~$A_{(m)}$ of the free-surface sides
in each element~$m$. Because all fluidic stresses are constrained to have only
normal components, we are free to use additional tangential force components for
keeping the boundary mesh as regular as possible. This can be done during the
assembly of the system matrices by weighting the surface tension by the
element's side length~$A_{(m)}$, divided by the average length~$\langle
A_{(m)}\rangle$ of all element sides contributing to the free surface. Of
course, this weighting factor becomes ineffective if all sides have equal
length. Any length difference of adjacent sides causes an additional force that
tries to equalize them. The tension forces for each element side are then
equivalent to a first variation of the functional~$\tension A_{(m)}^2 /
(2\langle A_{(m)}\rangle)$, which describes a rubber band with Hookean forces.
Instead of~$\variat F[\variat\bft]$ from Eq.~\eqnref{eq:var-t} we thus assemble
on each element
\begin{equation}
  \frac{\tension}{2\langle A_{(m)}\rangle}\variat(A_{(m)}^2)[\variat\bft]
   = \tension\frac{A_{(m)}}{\langle A_{(m)}\rangle}\variat A_{(m)}[\variat\bft].
\end{equation}
The second variations of $A_{(m)}$ and $A_{(m)}^2/2$ are not proportional to
each other,
\begin{multline}
  \label{eq:stab-secvar}
  \frac{\tension}{2\langle A_{(m)}\rangle}
  \variat^2(A_{(m)}^2)[\variat\bft,\variat\bft]
   = \\
     \tension\frac{A_{(m)}}{\langle A_{(m)}\rangle}
     \variat^2 A_{(m)}[\variat\bft, \variat\bft]
   + \tension \frac{1}{\langle A_{(m)}\rangle}
     \left(\variat A_{(m)}[\variat\bft]\right)^2
\end{multline}
In the implementation we therefore took only the first term on the right-hand
side of Eq.~\eqnref{eq:stab-secvar}. In this sense we did not strictly implement
the behavior of a rubber band, but yet a stabilized version of the free-surface
tension terms. After convergence all boundary sides of the mesh representing the
free surface have equal lengths and the extra terms $A_{(m)}/\langle
A_{(m)}\rangle$ do not change the behavior of the free surface.

\section{Summary of the algorithm}
\label{sec:summary}

Here, we provide a short overview of the complete algorithm. The
required steps are as follows:

\bigskip
\begingroup
\raggedright
\begin{enumerate}
  \renewcommand{\theenumi}{\arabic{enumi}}%
  \item Choose an initial mesh and initial ambient pressure~$p_0$.
  \item Until convergence repeat the following steps:
  \medskip
  \begin{enumerate}
    \makeatletter
    \renewcommand{\theenumii}{{\normalfont\itshape\@alph\c@enumii\/}}%
    \makeatother
    \item Smooth the inner mesh if it is too distorted.
    \item Repeatedly solve the fluidic system for $p$, $u$~and $v$, until the
    slip boundary condition is established.
    \item Subtract the average from $p$.
    \item Solve the geometry system for the new boundary. At the same time
    search for the value of $p_0$ that keeps the volume unchanged.
    \item Set the mesh boundary nodes to the parameterization values of the
    geometry system.
  \end{enumerate}
\end{enumerate}
\endgroup
\bigskip

The fluidic system is assembled according to
Eqs.~\eqnref{eq:fluidicmatrix}--\eqnref{eq:fluidicintegrals-end} with
constraints that account for the proper boundary conditions. To give the full
algorithm at this point, we summarize also the terms of the geometry system.
\begin{widetext}
The update Eq.~\eqnref{eq:geometricmatrix} is written as
\begin{equation}
  \label{eq:newton}
  \left(\begin{array}{cc}
    K_{rr} &
    K_{rs} \\
    K_{sr} &
    K_{ss}
  \end{array}\right)
  \left(\begin{array}{c}
    \vecr^\text{\,(new)} \\
    \vecs^\text{\,(new)}
  \end{array}\right)
  =  - \tau
  \left(\begin{array}{c}
    L_r \\
    L_s
  \end{array}\right) +
  \left(\begin{array}{cc}
    K_{rr} &
    K_{rs} \\
    K_{sr} &
    K_{ss}
  \end{array}\right)
  \left(\begin{array}{c}
    \vecr^\text{\,(old)} \\
    \vecs^\text{\,(old)}
  \end{array}\right)
\end{equation}
with entries that are assembled per element~$m$,
\newcommand{\integral}{\mkern-5mu\int\limits_{A_{(m)}}\mkern-7mu}
\begin{align}
  \bigl[L_r^{(m)}\bigr]_{d}
   &= \integral \chi_d \; (\bfe_x\scalarmult\sigma\scalarmult\bfN) \:dA
    + \frac{\tension A_{(m)}}{\langle A_{(m)}\rangle}
      \integral (\bfnabla\chi_d\scalarmult\bfT) (\bfe_x\scalarmult\bfT) \:dA\\
  \bigl[K_{rr}^{(m)}\bigr]_{de}
   &= \begin{aligned}[t]
      - \integral \chi_d \chi_e (\bfe_x\scalarmult\bfnabla p) (\bfe_x\scalarmult\bfN) \:dA
      + \integral \chi_d \; (\bfnabla\chi_e\scalarmult\bfT) \Bigl\{
           (\bfe_x\scalarmult\sigma\bfN)(\bfe_x\scalarmult\bfT)
         - (\bfe_x\scalarmult\sigma\bfT)(\bfe_x\scalarmult\bfN)
        \Bigr\} \:dA&{} \\[-2ex]
     {} + \frac{\tension A_{(m)}}{\langle A_{(m)}\rangle}
        \integral (\bfnabla\chi_d\scalarmult\bfT)(\bfnabla\chi_e\scalarmult\bfT) \:dA&{}
      \end{aligned}\\
  \bigl[K_{rs}^{(m)}\bigr]_{de}
   &= - \integral
       \chi_d \chi_e (\bfe_x\scalarmult\bfnabla p) (\bfe_y\scalarmult\bfN)
       \:dA
     + \integral
       \chi_d \; (\bfnabla\chi_e\scalarmult\bfT) \Bigl\{
          (\bfe_x\scalarmult\sigma\bfN)(\bfe_y\scalarmult\bfT)
        - (\bfe_x\scalarmult\sigma\bfT)(\bfe_y\scalarmult\bfN)
       \Bigr\} \:dA.
\end{align}
The remaining entries can be obtained by permutations of $x$~and $y$ together
with $r$~and $s$. Again, constraints have been used to keep the contact-lines
pinned. \end{widetext}

\section{Numerical experiments}
\label{sec:numerics}%

We performed all our test cases for a two-dimensional fluid. The programs were
written using the open-source C++~library \textit{libmesh}~\cite{libmesh} which
allows to change the elements' geometry in a user's routine and provides a
powerful constraint method.

\subsection{The instability of a ``direct explicit update'' algorithm}
\label{sec:explicit}%

\begin{figure}%
  \centerline{\includegraphics{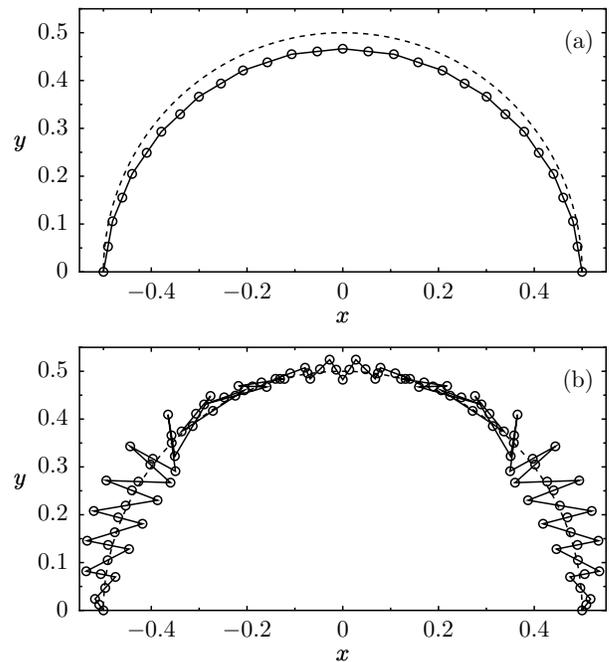}}%
  \caption{The stability of a direct explicit update algorithm strongly depends
  on the ratio of the step-size~$\tau$ and the element size. We have used the
  same $\tau=0.05$ for two different numbers of approximating FEs (first-order).
  A prescribed homogeneous pressure $p_0=2$ is applied which bends the free
  surface into a half-circle with radius~$1/2$ (with $\tension=1$). Panel~(a)
  depicts the converged result for 14~FEs after more than 500~steps. Panel~(b)
  shows a mesh with 52~FEs after only 12~steps. For this combination of
  step-size and element size the direct update algorithm is unstable, and the
  mesh was completely destroyed after a few more steps. Using second-order FEs
  the instability was similar. In both panels the dashed half-circles indicate
  both, the exact solution and the initial geometry.}%
  \label{fig:explicit}%
\end{figure}

In our first numerical example we do not use Newton's method with the
update-rule~\eqnref{eq:newton}, but instead with the direct and explicit
update
\begin{equation}
  \label{eq:direct-update}
  \left(\begin{array}{c}
    \vecr^\text{\,(new)} \\
    \vecs^\text{\,(new)}
  \end{array}\right)
  =  - \tau
  \left(\begin{array}{c}
    L_r \\
    L_s
  \end{array}\right) +
  \left(\begin{array}{c}
    \vecr^\text{\,(old)} \\
    \vecs^\text{\,(old)}
  \end{array}\right).
\end{equation}
The stability of this update rule delicately depends on the step-size
parameter~$\tau$. The allowed range of~$\tau$ strongly depends on the size of
the elements, the curvature, etc. Figure~\figref{fig:explicit} depicts the most
simple situation where a homogeneous pressure field deforms the boundary into a
circular arc with radius $R=-1/\curv = p_0/\tension$. The
update-rule~\eqnref{eq:direct-update} is stable for 14~FEs and a given step-size
while it is unstable for the same step-size with 52~FEs.

\subsection{Testing the accuracy of the Newton algorithm}
\label{sec:accuracy}%

In order to confirm the accuracy of the curvature approximation we have tested
two cases that can be solved analytically. Similar to the calculation in
Appendix~\secref{sec:static}, a prescribed pressure determines the free
surface's shape. Then, the approximation in Eq.~\eqnref{eq:stressvariat} becomes
exact and simplifies to
\begin{equation}
 \label{eq:stressvariat-exact}
 \variat \sigma_{ij} = -g_{ij} p_{,k}\variat t^k.
\end{equation}
Thus, all possible approximation errors must be due to the discretization of the
curvature.

\begin{figure}%
  \centerline{\includegraphics{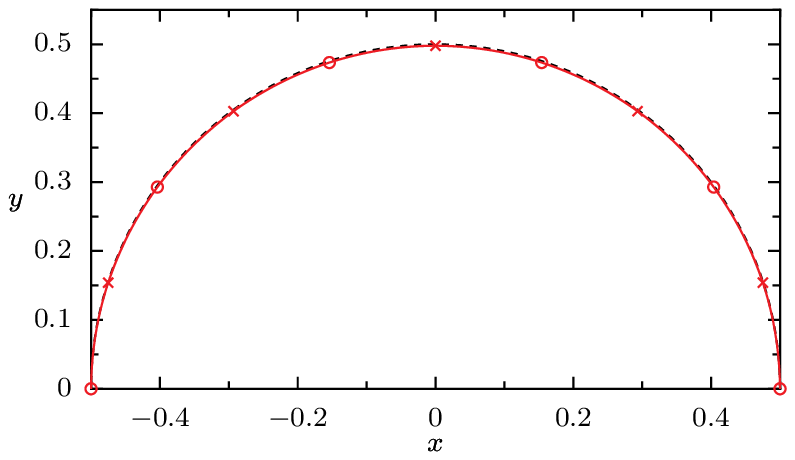}}\par\medskip\noindent
  \centerline{\includegraphics{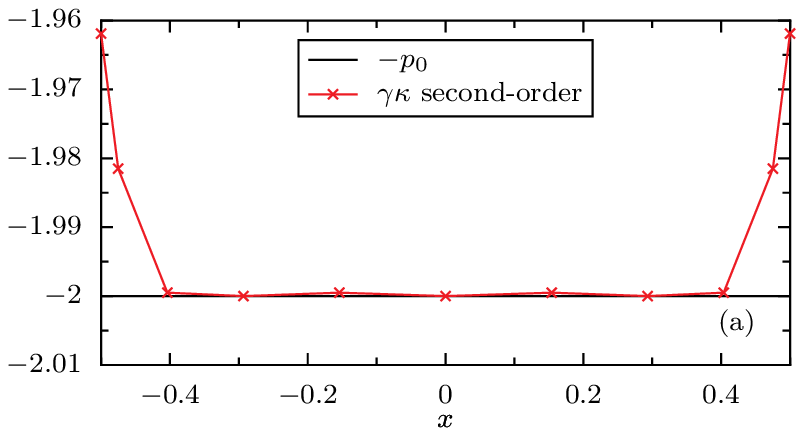}}%
  \caption{A prescribed homogeneous pressure $p_0=2$ bends the free surface into
  a half-circle with radius~$1/2$ (with $\tension=1$). Panel (a) presents the
  approximation with only five~second-order FEs. The boundary nodes are
  indicated by circles (every second is a second-order node). The exact solution
  is indicated by the dashed half-circle, while the starting geometry was the
  straight connection between the fixed endpoints. Good convergence was reached
  after 100~iterations with a step-size parameter $\tau=1$. The topmost node
  misses its exact position only by a relative error of only~$8.6\times10^{-6}$.
  This is also the error of the overall curvature approximation. In panel (b) an
  estimate of the curvature~$\curv$ was obtained by reconstructing the normal
  vectors and their change along the surface from the boundary shape of the FEs.
  This estimate compares very well with its expected value of $-2.0$. The
  outliers near the contact points are artefacts due to the reconstruction of the
  normal vectors.}%
  \label{fig:circle}%
\end{figure}

Figure~\figref{fig:circle}a depicts the most simple situation where a
homogeneous pressure field deforms the boundary into a circular arc, as in the
previous example. The free surface shape is approximated by the sides of
5~second-order~FEs. In dimensionless units the surface tension is~$\tension=1$,
and the prescribed pressure $p_0=2$ produces as the exact solution a circle with
radius $R=1/2$. The relative error of the numerically resulting radius, and thus
also of the curvature, is about $8.6\times 10^{-6}$. This value has been
obtained from the position of the topmost node. An alternative approach for
calculating the approximation error is visualized in Fig.~\figref{fig:circle}b.
We calculated the normal vectors at each node from the resulting
finite-element's side. Due to the elements being second-order we got a single
normal vector for second-order nodes. At vertices, where two elements meet and
where the surface parameterization is not smooth, we averaged the two normal
vectors. The curvature estimate at a node in Fig.~\figref{fig:circle}b is then
given by the curvature radius of a circle that connects the two neighbors of the
specific node, given their appropriate normal vector. Thus, we explicitly
reconstructed the curvature from the change of the normal vector along the
surface. It is clear by construction that the normal vector of the contact-nodes
cannot be correctly estimated. This causes the four outliers in
Fig.~\figref{fig:circle}b. All other nodes fit well.

A comparison with Fig.~\figref{fig:explicit}a, where the result of a
first-order approximation can be seen, makes clear that it is crucial to use a
second-order parameterization. The relative error of the curvature in
Fig.~\figref{fig:explicit}a is $5.0\times 10^{-3}$, three magnitudes larger
than in Fig.~\figref{fig:circle}a.
\begin{figure}%
  \centerline{\includegraphics{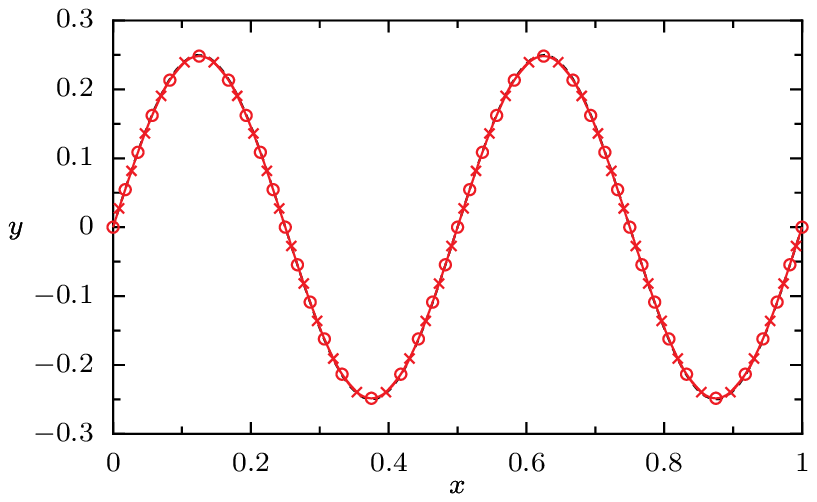}}\par\medskip\noindent
  \centerline{\includegraphics{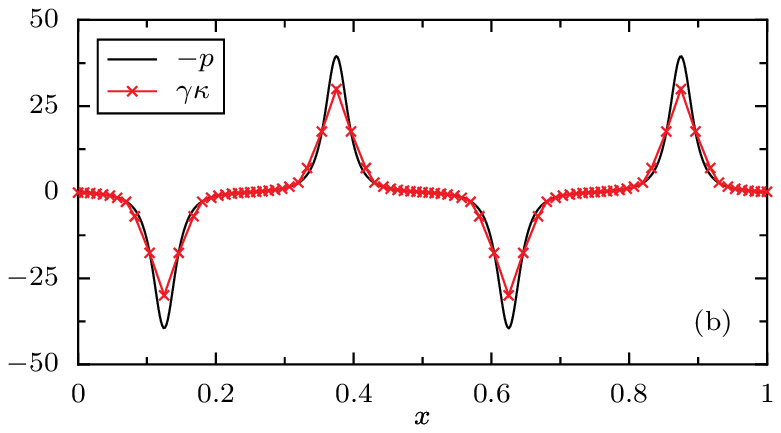}}%
  \caption{In panel (a) the expected sinusoidal boundary shape $y = h(x) = 0.25
  \sin(4\pi x)$ with surface tension $\tension=1$ is well recovered by
  40~second-order FEs. The shape is generated by the prescribed pressure of
  Eq.~\eqnref{eq:sinuspressure}. As in Fig.~\figref{fig:circle} the nodes are
  indicated by circles, the exact solution by the dashed curve and the initial
  geometry was the straight connection between the fixed endpoints of the
  surface. Good convergence was reached after 60~iterations with a step-size
  parameter $\tau=1$. The nodes' position at the maxima is off by a relative
  error of~$6.7\times 10^{-3}$ ($2.3\times 10^{-3}$ for 80~FEs and $2.5\times
  10^{-4}$ for 120~FEs, not shown). The elements' side-lengths vary only by
  $\pm0.007\%$. This small deviation demonstrates that the mesh regularization
  method does not influence the final behavior of the free boundary. The
  approximation quality is dramatically worse for 40~first-order FEs, yielding
  an estimated error of~$2.0\times10^{-1}$ (not shown). In Panel (b) the applied
  pressure is compared with curvature estimated by a reconstruction of the
  normal vectors. The large deviations at the extrema do not affect the overall
  approximation of the sinusoidal shape.}%
  \label{fig:sinus}%
\end{figure}%

In the next accuracy test, depicted in Fig.~\figref{fig:sinus}, the pressure is
still prescribed, but it varies in space. As above, we apply a pressure for
which the resulting boundary shape is known. Figure~\figref{fig:sinus}a
illustrates the approximation of a sinusoidal boundary height function $y = h(x)
= \alpha\sin(\beta x)$ that is caused by the corresponding pressure field
\begin{equation}
  \label{eq:sinuspressure}
  p(x,y)
  = -\tension \kappa(x)
  = \frac{1}{\tension} \frac{\alpha\beta^2\sin(\beta x)}%
    {[1+\alpha^2\beta^2\cos^2(\beta x)]^{3/2}}.
\end{equation}
Again, the approximation in Eq.~\eqnref{eq:stressvariat} becomes exact, and we
expect the same discretization errors as in the previous example. The
curvature's relative error is larger than in the previous example because the
curvature is bigger compared to the number of nodes. Nevertheless, the error is
still small enough to return the expected boundary shape within reasonable
accuracy. It decreases with the number of approximating elements. If a
first-order approximation is used it is much larger, maybe intolerably large.

Concerning the discretization errors of the curvature the accuracy test in
Fig.~\figref{fig:sinus} covers already the general case. According to the
construction of the algorithm the flow exerts stress on the boundary only in
normal direction. It makes no difference whether this stress is of viscous
nature or due to a pressure difference.

\subsection{A deformed micro-droplet}
\begin{figure}%
  \centerline{\includegraphics{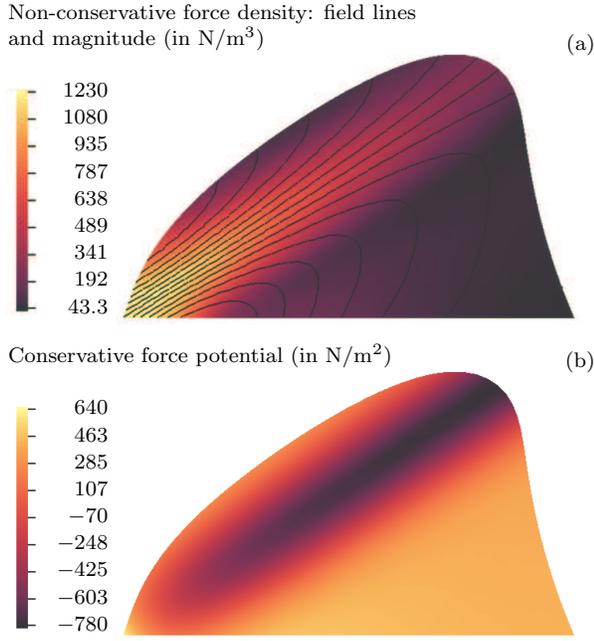}}\par\medskip\noindent
  \centerline{\includegraphics{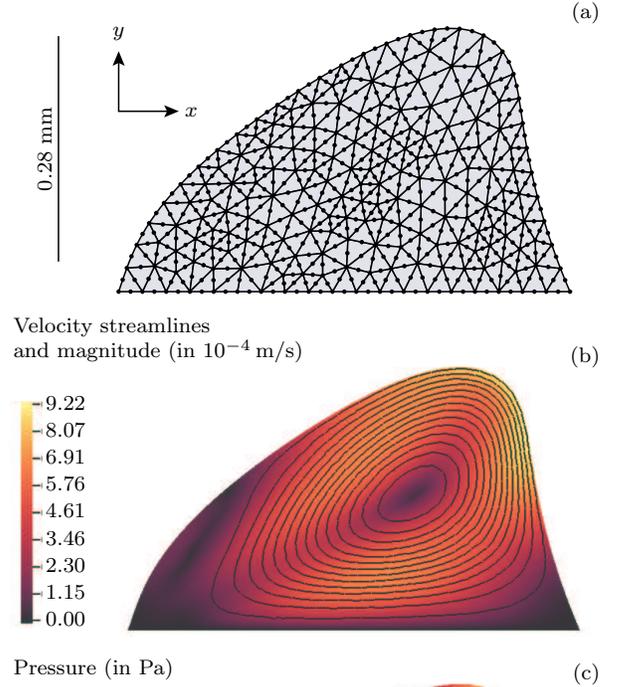}}%
  \caption{The force density that models the effect of the SAW in the droplet
  given in Fig.~\figref{fig:droplet}. Panel (a) depicts the non-conservative
  part that causes the flow; (b) shows the potential of the conservative part
  that contributes only to the pressure. The same non-conservative force density
  has also been used in Fig.~\figref{fig:weakdroplet}.}%
  \label{fig:driving}%
\end{figure}%
\begin{figure}%
  \centerline{\includegraphics{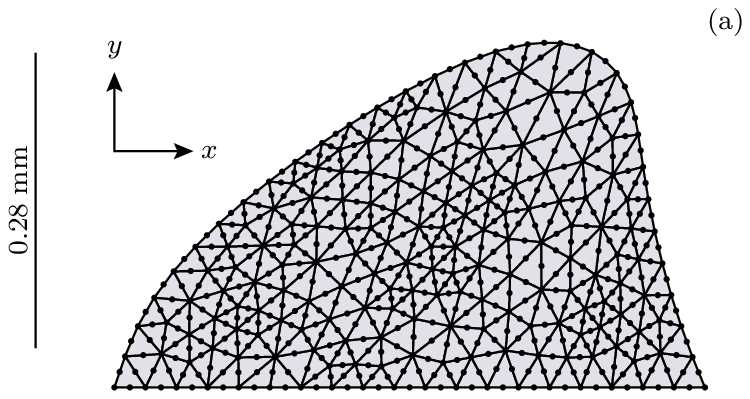}}\par\medskip\noindent
  \centerline{\includegraphics{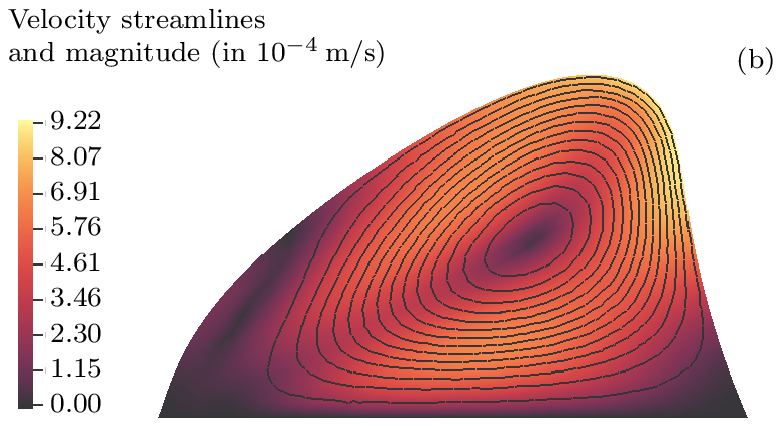}}\par\medskip\noindent
  \centerline{\includegraphics{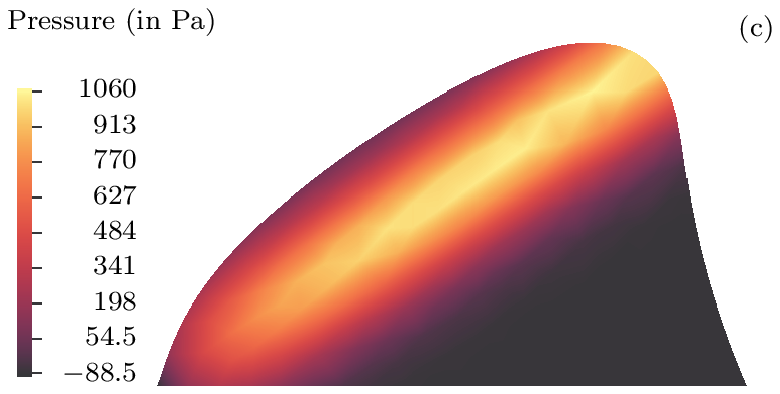}}\par\medskip\noindent
  \centerline{\includegraphics{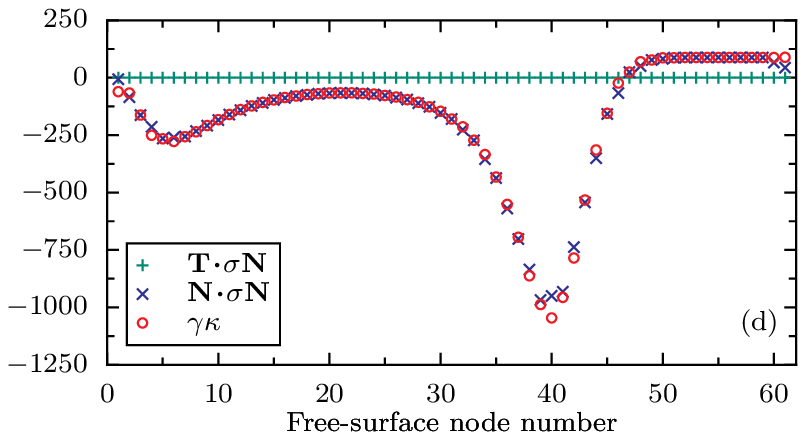}}%
  \caption{A deformed micro-droplet, sitting on a flat substrate with pinned
  contact points. The deformation is due to an internal pressure and viscous
  flow, both caused by the body force density illustrated in
  Fig.~\figref{fig:driving}. The material properties are those of water
  surrounded by air at room temperature. Its two-dimensional ``volume'' is that
  of the initial half-circle with radius 0.28\,mm. Panel (a) illustrates the
  computational grid, consisting of second-order elements. The side-lengths of
  the free-surface facets differ only by $4.5\times10^{-5}\%$. Panels (b)~and
  (c) depict the flow and the pressure, respectively. Note that the deformation
  is predominantly caused by the pressure which corresponds to the case that
  $\Capillary\ll\Bond$. Good convergence was reached after 7~iterations with a
  step-size parameter $\tau=0.5$.}%
  \label{fig:droplet}%
\end{figure}%
\begin{figure}%
  \centerline{\includegraphics{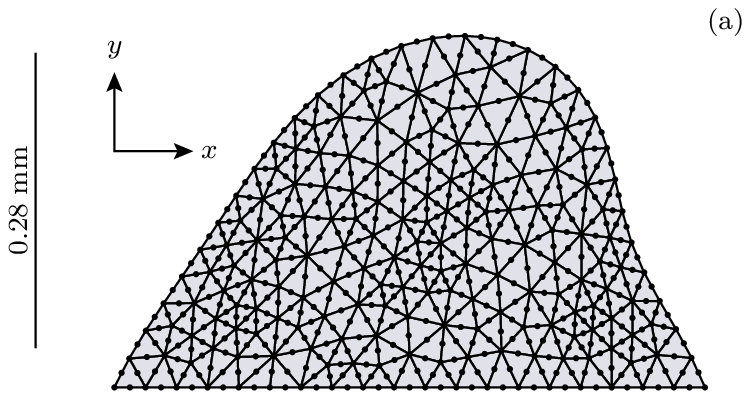}}\par\medskip\noindent
  \centerline{\includegraphics{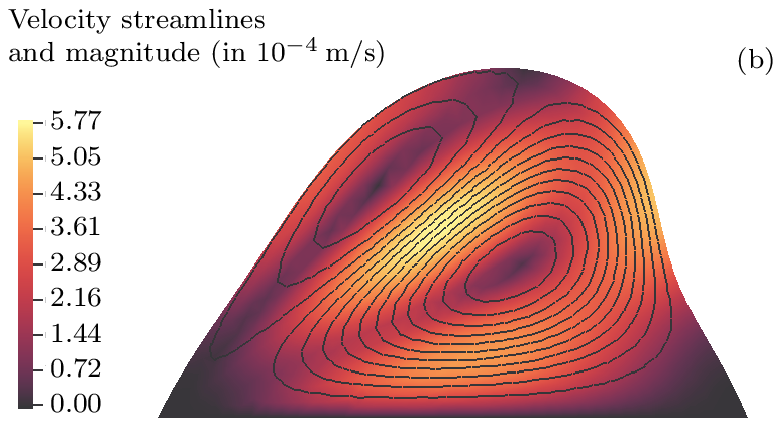}}\par\medskip\noindent
  \centerline{\includegraphics{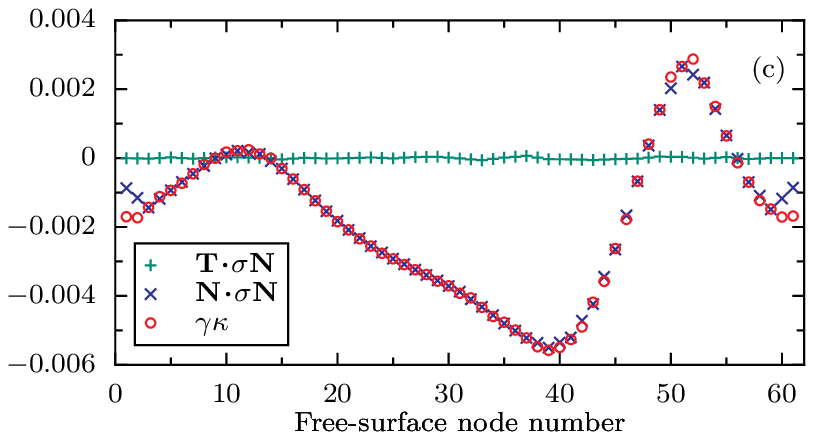}}%
  \caption{A similar micro-droplet as in Fig.~\figref{fig:droplet}, deformed
  only by the viscous stress at the boundary. The flow is driven by the
  non-conservative force density depicted in Fig.~\figref{fig:driving}a. The
  conservative part of the force vanishes such that the pressure is constant
  inside the droplet. In order to obtain a comparable deformation we have used a
  $10^5$~times smaller surface tension than that of a water--air interface. This
  corresponds to the case $\Bond\ll\Capillary$. Good convergence was
  achieved after 30~iterations with $\tau=0.1$.}%
  \label{fig:weakdroplet}%
\end{figure}%

In order to explicitly show that the quality of the curvature discretization
does not depend on the origin of the applied normal stress we like to return to
the introductory motivation for the present work. The previous examples were
analytically solvable. The form and internal streaming of micro-droplets,
however, cannot be determined analytically.

In the experiment the internal flow is agitated by a surface-acoustic wave~(SAW)
due to the \emph{acoustic streaming effect.}\cite{Nyborg65} Because the very
details of the SAW's impact are not known, we here model it by a body force that
is active in the fluid only, as depicted in Fig.~\figref{fig:driving}. The force
is concentrated in a narrow channel starting at the left contact point where the
SAW hits the fluid and continuing into the fluid. It essentially carries the
fluid along this channel, from the entry point of the SAW into the droplet,
giving rise also to a back-flow.\cite{WixGut} Additionally, the force has a
strong conservative portion that is balanced by the pressure in the fluid.

The resulting stationary droplet shape and the internal velocity and pressure
fields are presented in Fig.~\figref{fig:droplet}. The initial shape was a
half-circle with the same two-dimensional volume. The deformed boundary consists
of two regions, one with negative curvature (as the initial half circle) and
another one at the right flank of the droplet with positive curvature. The
material properties are those of water and air at room temperature, i.~e.\
$\viscosity=10^{-3}$\,kg/ms and $\tension=72.8\times10^{-3}$\,N/m. The --
admittedly strange -- deformation of the droplet qualitatively agrees with the
experimentally observed jumping droplet in Fig.~4 of the publication by
Wixforth~\etal.\cite{WixStrGau04} The deformation is due to the large
conservative contribution of the driving force and the resulting pressure. The
viscous forces for the given velocities are far too weak to lead to a
substantial deformation of the free surface. The capillary number for the
illustrated flow is $\Capillary\approx10^{-5}$, the Bond number is around one.
Although the free surface is significantly deformed, its discretization by
finite-element sides is as regular as possible. Their lengths vary only
by~$4.5\times10^{-5}\%$. This guarantees that the behavior of the boundary is
indeed that of a free surface and is not disturbed by the automatic
regularization technique described in Sec.~\secref{sec:rubber}.
Figure~\figref{fig:droplet}d quantifies the normal stress condition. For each
node we integrated the normal and the tangential component of the normal stress,
weighted with the corresponding ansatz function of the node. The tangential
component vanishes perfectly. The normal component coincides well with the
reconstruction estimate of the curvature as in the previous examples. Thus, the
free-surface boundary condition is indeed satisfied.

In order to prove that our algorithm can likewise produce stable results in the
parameter regime $\Bond\ll\Capillary$ we consider a droplet that is deformed
only by viscous stress at the boundary. In Fig.~\figref{fig:weakdroplet} we have
used the same non-conservative force that is visualized in
Fig.~\figref{fig:driving}, but we omitted the conservative part. Thus, the
pressure was constant and $\Bond=0$. With the same water--air interface tension
as in the previous example the droplet would hardly be deformed. To obtain a
comparable deformation as in the previous case together with $\Capillary\approx
1$ we took an artificial $10^5$~times smaller surface tension. In this example
the stress that deforms the free surface depends much stronger on the shape of
the surface itself. Thus, the approximation in Eq.~\eqnref{eq:stressvariat}
becomes questionable. It was necessary to reduce the step-size parameter $\tau$
to a smaller value than in the previous examples.

\section{Summary and Outlook}
Within this work we presented a weak formulation of free-surface boundary
problems in arbitrary coordinate systems. The steps of the derivation are
physically and mathematically founded using variational techniques for the
Stokes equations and the differential geometry of the surface. We found that the
applicability of different numerical treatments for the curvature terms depend
strongly on the scales of the system. Our method is designed for Bond and
capillary numbers assuming values from zero up to unity. Which one is larger
plays no role.

A decisive benefit of our method is the automatic control of mesh regularity at
a free surface. Many algorithms implementing the weak form of the free-surface
boundary condition encounter intrinsic instabilities of the boundary mesh.
Often, it is therefore necessary to create a completely new mesh after several
iteration steps. Our formulation includes a smooth transition to the behavior of
a rubber band when the boundary mesh becomes distorted. This leads to an
inherent regularization of the mesh without affecting the behavior of the free
surface.

As another important result we find that for physical reasons the geometry
variables for the parameterization of the free surface should be approximated on
the same level of accuracy as the velocity variables. This substantiates
numerical observations reported by B\"ansch.\cite{Bansch98}

The quality of our numerical approach is tested by two analytically solvable
examples. We explicitly plot the curvature of the free surface and the stress
that causes the deformation. This confirms that the free surface boundary
condition is indeed satisfied, not only in the weak sense which is implemented
but even leads to a reliable reconstruction of the curvature by the normal
vectors of the finite element's sides. Two further examples illustrate that the
ratio of capillary number and Bond number has only a weak influence on the
stability of the algorithm.

The presented covariant formulation opens the possibility to utilize the
powerful differential geometric description of free surfaces in finite-element
implementations of the Stokes equations. It thus provides a natural approach to
treat surfaces and interfaces with a richer behavior such as lipid vesicles
containing bending stiffness, area constraints, and much more. Many potential
applications can be found in the literature on lipid vesicle geometry, where
more complicated expressions for the surface's free energy contribution are in
use.\cite{Guven04,CapGuvSan03,Seifert97}

Extensions of the presented approach towards moving contact-lines, towards
time-dependent flows and a three-dimensional implementation are possible. There
are still some hurdles to be overcome that can be clearly seen in our
derivation. One of them is the principally unknown mutual dependence of the
stress tensor and the surface parameterization where we had to introduce the
approximation~\eqnref{eq:stressvariat}. Another one is the understanding of the
numeric instabilities caused by the last integral in second variation of the
surface's free energy (see Eq.~\eqnref{eq:var-tt}).

These extensions would also provide a solid basis for the theoretical
understanding of particle transport in surface-acoustic-wave-driven
flows.\cite{SriStrSch05,StrSchBei04,StrFroGut,KosSchTalHae05}

\acknowledgments 
We gratefully acknowledge our experimental partners in the group of Prof.\
Achim~Wixforth, Univ.~Augsburg, and the developers of the \emph{libmesh}
project. This work was supported by the \emph{Deut\-sche
Forsch\-ungs\-ge\-mein\-schaft} (DFG) via grant 1517/25-1, SFB 486 and the
\emph{Gradu\-ier\-ten\-kol\-leg: Nicht\-line\-are Pro\-ble\-me in Ana\-lysis,
Geo\-me\-trie und Physik}.

\appendix

\section{A static droplet in Cartesian coordinates}
\label{sec:static}%
The aim of this appendix is to recall the variational techniques in a simple
three-dimensional Cartesian setup before going to arbitrary coordinates in
appendix~\secref{sec:diffgeom-var}. The argumentation is similar to the one
given by~Cuvelier.\cite{CuvSch90}

We describe the special case where the fluid's two-dimensional free surface~$A$
can be described by a height function
\begin{equation}
  A\colon z = h(x,y)
\end{equation}
which is non-zero over a certain region $(x,y)\in E$. The Stokes equations for
the static situation with a conservative force $f_i = -\Phi_{,i}$ reduce to
\begin{equation}
  \label{eq:trivialstokes}
  0 = - p_{,i} - \Phi_{,i}
\end{equation}
with the solution~$p(\bfx) = p_0 - \Phi(\bfx)$. The undetermined homogeneous
term~$p_0$ will be identified as the Lagrange multiplier for the
constraint of constant volume $V = \int_E dx\,dy\, h(x,y)$.

The free energy of the system consists of the surface-integral of the
constant surface tension and the volume-integral of the potential
\begin{equation}
  F = \tension \int_A dA + \int_V \Phi\, dV
    = \int_E \calF(x,y) \,dx\,dy
\end{equation}
with
\begin{multline}
  \calF(x,y) =
       \tension \sqrt{1 + (\partial_x h(x,y))^2 + (\partial_y h(x,y))^2} \\
       + \int_0^{\hbox to 1.5em{$\scriptstyle h(x,y)$\hss}} \Phi(x,y,z)\,dz \:.
\end{multline}
The Euler-Lagrange equation for finding the extremal~$F$ by varying~$h$ is then
\begin{align}
  0 &= \frac{\formal \calF}{\formal h}
     - \frac{\partial}{\partial x}\frac{\formal \calF}{\formal(\partial_x h)}
     - \frac{\partial}{\partial y}\frac{\formal \calF}{\formal(\partial_y h)} \\
    &= \Phi(x,y,h(x,y)) - \tension\kappa(x,y)
\end{align}
where $\kappa$~is the curvature of~$A$, given by
\begin{multline}
  \kappa(x,y)
   = \frac{\partial}{\partial x} \frac{\partial_x h(x,y)}%
          {\bigl[1 + (\partial_x h)^2 + (\partial_y h)^2 \bigr]^{1/2}} \\
   + \frac{\partial}{\partial y} \frac{\partial_y h(x,y)}%
          {\bigl[1 + (\partial_x h)^2 + (\partial_y h)^2 \bigr]^{1/2}}.
\end{multline}
Because the pressure is given by the potential, the Euler-Lagrange equation is
equivalent to the free-surface boundary condition for a static fluid,
\begin{equation}
  -p(x,y,h(x,y)) + p_0 = \tension\kappa(x,y).
\end{equation}
At this point it is easy to see that~$p_0$ plays the role of a Lagrange
multiplier for a volume constraint. Adding the term
\begin{equation}
  \lambda V = \lambda \int_E dx\,dy\, h(x,y)
\end{equation}
to~$F$ gives an additional constant~$\lambda$ in the Euler-Lagrange equation,
just as the pressure offset~$p_0$. Because $p_0$~is yet undetermined we may
identify it with~$\lambda$.

\section{Variational calculus for the surface's parameterization}
\label{sec:diffgeom-var}%
In order to prove equality~\eqnref{eq:var-t} we express the change of the
surface's free energy functional~\eqnref{eq:func-tens} by the change of the
Jacobi determinant~$\sqrt{a}$ of the surface parameterization. With the
infinitesimal surface area $dA = \sqrt{a}\,d\bfnu$ the variation of the surface
free energy becomes
\begin{multline}
  \label{eq:var-t-app}
  \variat F[\variat\bft]
    = \variat\left( \int_A \tension\,dA\right)
    = \int_{E} \tension\: \variat\sqrt{a} \:d\bfnu \\
    = \int_{E} \tension\:
      \frac{\formal \sqrt{a}}{\formal t^i_{,\alpha}} \variat
      t^i_{,\alpha}\:d\bfnu.
\end{multline}
The dependence of~$\sqrt{a}$ on the tangent vectors follows from its definition
as the determinant of the surface metric. For a two-dimensional surface it reads
\begin{multline}\label{eq:2Da}
  a
  = \left|\begin{array}{cc}
      a_{11} & a_{12}\\ a_{21} & a_{22}
    \end{array}\right|
  = \frac{1}{2} \epsilon^{\alpha\gamma} \epsilon^{\beta\delta}
    a_{\alpha\beta} a_{\gamma\delta} \\
  = \frac{1}{2} \epsilon^{\alpha\gamma} \epsilon^{\beta\delta}
    g_{ij} g_{kl}
    t^i_{,\alpha} t^j_{,\beta} t^k_{,\gamma} t^l_{,\delta}
\end{multline}
where $\epsilon^{\alpha\beta}$ is the permutation symbol in two dimensions,
\begin{equation}
  \epsilon^{\alpha\beta} = \left\{\begin{array}{rl}
     0 & \alpha=\beta \\
    +1 & \alpha=1,\quad \beta=2\\
    -1 & \alpha=2,\quad \beta=1
  \end{array}\right.
\end{equation}
which is a relative surface tensor with weight $+1$. The absolute tensor
results~as
\begin{equation}
  \varepsilon^{\alpha\beta} = \frac{\epsilon^{\alpha\beta}}{\sqrt{a}}.
\end{equation}
This is analogous to the completely antisymmetric tensor in three dimensions,
described in detail by~Aris.\cite{Aris89} With the antisymmetric tensor we
obtain the inverse surface metric as
\begin{equation}
  a^{\alpha\beta} =
  \varepsilon^{\alpha\gamma} \varepsilon^{\beta\delta} a_{\gamma\delta}.
\end{equation}
A formal derivative of~\eqnref{eq:2Da} yields
\begin{align}
  \frac{\formal a}{\formal t^i_{,\alpha}}
  &= 2 g_{ij} t^j_{,\beta}
     \varepsilon^{\alpha\gamma} \varepsilon^{\beta\delta} a_{\gamma\delta}
   = 2 a\:  g_{ij} a^{\alpha\beta} t^j_{,\beta}\quad\text{and}\\
  \frac{\formal \sqrt{a}}{\formal t^i_{,\alpha}}
  &= \frac{1}{2\sqrt{a}} \frac{\formal a}{\formal t^i_{,\alpha}}
   = \sqrt{a}\: g_{ij} a^{\alpha\beta} t^j_{,\beta}
\end{align}
which can be inserted into~\eqnref{eq:var-t-app} to give the desired
result~\eqnref{eq:var-t}.

For a one-dimensional curve in two-dimensional space the same formula can be
derived, but the notation may be somewhat confusing. Summation over the single
surface index makes no sense, nevertheless, we still have to distinguish between
co- and contravariant relative tensors, i.~e.,
\begin{gather}
  a = a_{11} = g_{ij} t^i_{,1} t^j_{,1} \\
  a^{11} = 1/a_{11} \quad\text{because}\quad
  a^{11}a_{11} = a^{\alpha\beta}a_{\alpha\beta} = 1.
\end{gather}
The formal derivative then becomes
\begin{equation}
  \frac{\formal \sqrt{a}}{\formal t^i_{,1}}
  = \frac{1}{2\sqrt{a}} 2g_{ij} t^j_{,1}
  = \sqrt{a} g_{ij} t^j_{,1} \frac{1}{a_{11}}
  = \sqrt{a} g_{ij} t^j_{,1} a^{11},
\end{equation}
which completes the result for the one-dimensional surface.


\section{Integration by parts of the tension forces}
\label{sec:curvature}

In order to see that Eq.~\eqnref{eq:var-t-partint} follows from
Eq.~\eqnref{eq:var-t} we remove the surface covariant derivative
from~$\variat t^i_{,\alpha}$ by an integration by parts and obtain
\begin{multline}
  \label{eq:dDdt_pi}
  \variat F[\variat\bft]
   = - \int_A \tension a^{\alpha\beta} t^i_{,\alpha\beta} g_{ij} \variat t^j
     - \int_A \tension_{,\beta} a^{\alpha\beta} t^i_{,\alpha} g_{ij} \variat t^j
     \\
     + \oint_{\partial A} \tension \nu_{\beta} a^{\alpha\beta} t^i_{,\alpha} g_{ij}
     \variat t^j
\end{multline}
where the covariant surface vector~$\nu_{\beta}$ is tangential to~$A$ and normal
to~$\partial A$. We can express the surface-derivatives~$t^i_{,\alpha\beta}$ by
the tensor~$b_{\alpha\beta}$ of the second fundamental form of the surface from
Eq.~\eqnref{eq:secondfundf} (cf. to Aris,\cite{Aris89} p.~216),
\begin{equation}
  t^i_{,\alpha\beta} = b_{\alpha\beta} N^i,
\end{equation}
arriving at
\begin{equation}
  a^{\alpha\beta} t^i_{,\alpha\beta}
  = a^{\alpha\beta} b_{\alpha\beta} N^i
  = \curv N^i.
\end{equation}
We have used the definition of the curvature as the trace of the tensor of the
second fundamental form as in Eq.~\eqnref{eq:curvature}. For a two-dimensional
surface this is twice the mean curvature $\curv = 2H = a^{\alpha\beta}
b_{\alpha\beta}$, for a one-dimensional surface we have only one entry $\curv =
a^{11} b_{11}$.

As consistent with the standard literature,\cite{LanLif63,Aris89} the term $\variat
F$~from Eq.~\eqnref{eq:var-t} comprises a curvature term in normal
direction
\begin{equation}
  -\tension \curv \, N^i
\end{equation}
and a term accounting for the surface-gradient of~$\tension$. The space
vector
\begin{equation}
  \label{eq:tensiongrad}
  -t^i_{,\alpha} a^{\alpha\beta} \tension_{,\beta}
\end{equation}
is tangential to the surface. The third term on the right-hand side of
Eq.~\eqnref{eq:dDdt_pi}, which is an integral over the
contact-line~$\partial A$, vanishes because for a pinned droplet $\variat t^i =
0$~vanishes on the contact-line.

\section{Invoking constraints for the slip boundary condition}
\label{sec:slipconstraints}%
\begin{figure}%
  \centerline{\includegraphics{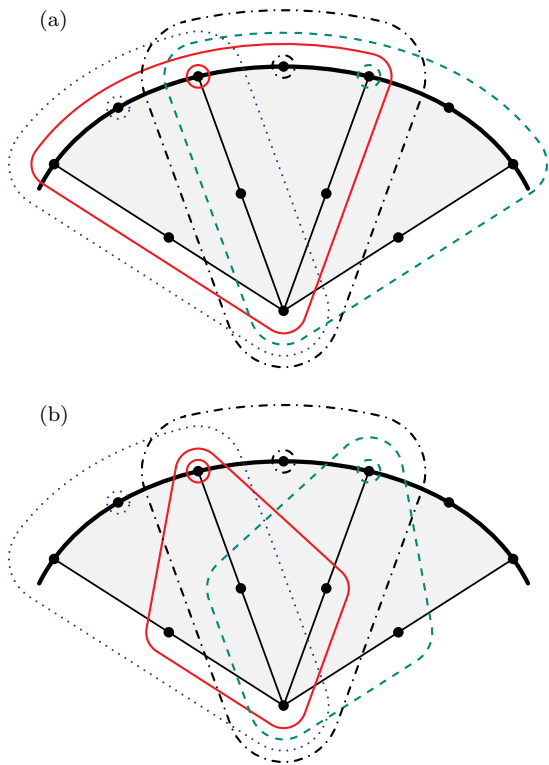}}%
  \caption{A sketch of the cross-dependencies among the DoFs located on three
  elements. The free surface is indicated by the thick curve. Constrained DoFs
  are surrounded by small circles. The nodes carrying the corresponding
  constraining DoFs are surrounded by curves drawn in the same style (solid and
  dashed for vertices; dotted and dash-dotted for second-order nodes). Panel~(a)
  depicts the full inter-dependencies while in~(b) the DoFs located at vertices
  depend only on DoFs located at inner nodes. By taking the values of the
  missing adjacent DoFs located on the free surface as inhomogeneities instead
  of constraints in~(b) constraints are decoupled. The correct constraint
  equations are then established after some iteration steps.}%
  \label{fig:slip}%
\end{figure}%

The tangential components of the free-surface boundary condition correspond to a
slip boundary condition. When this condition is expressed as a set of
constraints for the DoFs, we obtain one equation like~\eqnref{eq:free-bc-t-w}
per each DoF at the free surface. Because the derivatives of the ansatz
functions~$\phi$ from~\eqnref{eq:ansatzphi} generally do not vanish at proximate
nodes, the constraint equations contain non-vanishing weights for all DoFs that
are located on the same element. Therefore, the constraints for DoFs that are
connected to two adjacent elements create inter-dependencies of DoFs also on
other elements. This is illustrated in Fig.~\figref{fig:slip}a. As a result,
all DoFs on the free surface implicitly depend on each other. After the
constraints are re-sorted such that DoFs are constrained only in terms of
non-constrained ones, it turns out that the free-surface DoFs depend on all DoFs
in the element layer near the surface.

As a strategy to avoid this full dependency we replace the constraint equation of
type~\eqnref{eq:constraint} by
\begin{equation}
  \label{eq:slipconstraints}
  u_d = w_d + \sum_{e\in\Lambda_d} w_{de} u_e
            + \sum_{e\in\overline{\Lambda}_d} w_{de} u_e^\text{(old)},
\end{equation}
where the sums run over two complementary sets $\Lambda_d$~and
$\overline{\Lambda}_d$. The DoFs in~$\Lambda_d$ contribute to the constraint for
$u_d$ in the usual way, while those in~$\overline{\Lambda}_d$~have been
substituted by their old values~$u_e^\text{(old)}$ and thus contribute to the
inhomogeneity. There is some freedom in the choice, which of the participating
DoFs in one element are in~$\Lambda_d$ and which are taken
into~$\overline{\Lambda}_d$. We found that the combination illustrated in
Fig.~\figref{fig:slip}b works well: For the DoFs located at element vertices we
take the DoFs that belong to adjacent nodes on the free surface as
inhomogeneities; all other constraining DoFs are located at inner nodes and are
not constrained. The DoFs located at the second-order nodes on the free surface
acquire their full constraints. When all constrained DoFs are expressed by
non-constrained DoFs, then the resulting constraint equations will only contain
DoFs that are located at inner nodes of three adjacent elements. This presents a
sufficient decoupling of the constraint equations to yield an efficient
algorithm.

Although the boundary condition given by Eq.~\eqnref{eq:slipconstraints} is not
the correct one when the true velocity field has not yet been determined, it
still improves as the velocity field tends to the proper solution. Thus, there
is hope that the correct boundary condition is established by the successive use
of Eq.~\eqnref{eq:slipconstraints} using increasingly good values for the
values~$u_e^\text{(old)}$. In numerical experiments the scheme for splitting the
cross-dependencies as illustrated in Fig.~\figref{fig:slip}b turned out to be
the only one that works. In the examples of Figs.~\figref{fig:droplet} and
\figref{fig:weakdroplet}, it took about 20 iteration steps to establish the
correct boundary condition from scratch, and 5 iteration steps to re-establish
it after a change of the mesh. This could be readily observed because after the
first iteration step the velocity field exhibited oscillations at the boundary
nodes that ceased during iteration.


\end{document}